\documentclass[aps,physrev,twocolumn,superscriptaddress]{revtex4-2}

\usepackage{graphicx}
\usepackage{dcolumn}
\usepackage{bm}
\usepackage[export]{adjustbox}
\usepackage{multirow} 

\begin{document}

\title{A Neural Network-Based Search for Unmodeled Transients in LIGO-Virgo-KAGRA's Third Observing Run}

\author{Ryan Raikman}
\affiliation{Department of Physics, MIT, Cambridge, MA 02139, USA}
\author{Eric A. Moreno}
\affiliation{Department of Physics, MIT, Cambridge, MA 02139, USA}
\affiliation{LIGO Laboratory, 185 Albany St, MIT, Cambridge, MA 02139, USA}
\author{Katya Govorkova}
\affiliation{Department of Physics, MIT, Cambridge, MA 02139, USA}
\affiliation{LIGO Laboratory, 185 Albany St, MIT, Cambridge, MA 02139, USA}

\author{Siddharth Soni}
\affiliation{LIGO Laboratory, 185 Albany St, MIT, Cambridge, MA 02139, USA}
\author{Ethan Marx}
\affiliation{Department of Physics, MIT, Cambridge, MA 02139, USA}
\affiliation{LIGO Laboratory, 185 Albany St, MIT, Cambridge, MA 02139, USA}
\author{William Benoit}
\affiliation{School of Physics and Astronomy, University of Minnesota, Minneapolis, MN 55455, USA}
\author{Alec Gunny}
\affiliation{Department of Physics, MIT, Cambridge, MA 02139, USA}
\affiliation{LIGO Laboratory, 185 Albany St, MIT, Cambridge, MA 02139, USA}
\author{Deep Chatterjee}
\affiliation{LIGO Laboratory, 185 Albany St, MIT, Cambridge, MA 02139, USA}
\author{Christina Reissel}
\affiliation{Department of Physics, MIT, Cambridge, MA 02139, USA}
\author{Malina M. Desai}
\affiliation{LIGO Laboratory, 185 Albany St, MIT, Cambridge, MA 02139, USA}
\author{Rafia Omer}
\affiliation{School of Physics and Astronomy, University of Minnesota, Minneapolis, MN 55455, USA}
\author{Muhammed Saleem}
\affiliation{School of Physics and Astronomy, University of Minnesota, Minneapolis, MN 55455, USA}

\author{Philip Harris}
\affiliation{Department of Physics, MIT, Cambridge, MA 02139, USA}
\author{Erik Katsavounidis}
\affiliation{Department of Physics, MIT, Cambridge, MA 02139, USA}
\affiliation{LIGO Laboratory, 185 Albany St, MIT, Cambridge, MA 02139, USA}
\author{Michael W. Coughlin}
\affiliation{School of Physics and Astronomy, University of Minnesota, Minneapolis, MN 55455, USA}
\author{Dylan Rankin}
\affiliation{Department of Physics and Astronomy, University of Pennsylvania, Philadelphia, PA, 19104, USA}

\email{Contact author: rraikman@mit.edu}

\date{\today}

\begin{abstract}
This paper presents the results of a Neural Network~(NN)-based search for short-duration gravitational-wave transients in data from the third observing run of LIGO, Virgo, and KAGRA. The search targets unmodeled transients with durations of milliseconds to a few seconds in the 30--1500\,Hz frequency band, without assumptions about the incoming signal direction, polarization, or morphology. Using the Gravitational Wave Anomalous Knowledge (\texttt{GWAK}) method, three compact binary coalescences (CBCs) identified by existing pipelines are successfully detected, along with a range of detector glitches. The algorithm constructs a low-dimensional embedded space to capture the physical features of signals, enabling the detection of CBCs, detector glitches, and unmodeled transients. This study demonstrates \texttt{GWAK}’s ability to enhance gravitational-wave searches beyond the limits of existing pipelines, laying the groundwork for future detection strategies.
\end{abstract}


\maketitle

\section{\label{sec:introduction}Introduction}

Gravitational-wave (GW) astronomy has revolutionized the way we observe and understand our universe. It provides a completely new way to study astronophysical  phenomena~\cite{PhysRevLett.116.061102}. The detection of gravitational waves by LIGO-Virgo-KAGRA (LVK) observatories~\cite{Aasi_2015, Acernese_2015, Aso_2013} marks a significant achievement in physics and astronomy. As these observatories' sensitivity increases, so does the complexity and volume of data collected during observation runs. This complexity necessitates advanced analytical techniques to identify and categorize both known and novel astrophysical events embedded within the data.

The third observation run (O3) of LVK was particularly fruitful, yielding numerous detections exclusively from compact binary coalescences (CBCs) such as binary black holes (BBH), binary neutron stars (BNS), and neutron star-black hole (NSBH) mergers~\cite{LIGOScientific:2020ibl,LIGOScientific:2021usb,PhysRevX.9.031040}. The traditional approach to analyzing such GW events has predominantly relied on matched-filter~\cite{helstrom1968,Allen:2005fk} algorithms which utilize pre-defined templates to identify and detect known patterns of gravitational wave signals. However, these methods are inherently limited by their dependence on existing knowledge of signal characteristics, which can restrict their ability to detect anomalies or novel phenomena that do not conform to pre-established templates. To address this limitation, unmodeled search methods, such as Coherent WaveBurst (cWB)\cite{Klimenko:2008fu}, oLIB \cite{Lynch:2015yin}, PySTAMPAS\cite{Macquet:2021ibe}, and BayesWave~\cite{Cornish:2014kda}, operate independently of template-based paradigms and are capable of detecting transient gravitational-wave signals of unknown or poorly modeled origin. The results of such unmodeled searches during O3 are presented in the LVK results paper~\cite{Abbott_2021, KAGRA:2021bhs}.

Deep learning methods for GW detection have been extensively studied, primarily relying on supervised techniques that exploit neural network nonlinearity and ground-truth labels~\cite{Baker:2014eba,George:2016hay, Kapadia:2017fhb,George:2017pmj,Gabbard:2017lja,Miller:2019jtp,Jadhav:2020oyt,Huerta:2020xyq, 9956104, Beveridge:2023bxa, Marx:2024wjt, mcleod2024binaryneutronstarmerger}. These approaches are effective but are limited to events similar to those in the training data.

Autoencoders, a type of neural network, find wide application in GW tasks, compressing input data into a smaller latent space and then reconstructing it. Since they only require a single dataset to train, they present a flexible network architecture for many applications. Currently, they enable noise subtraction, waveform generation, waveform profile extraction post-detection, and glitch classification, serving as efficient tools~\cite{Ormiston:2020ele, Bacon:2022lsm, Saleem:2023hcm, Gunny:2021gne, PhysRevD.103.124051, Chatterjee_2021, , Sankarapandian:2021qun}. Explorations into unsupervised GW detection aim to broaden detection beyond signal templates and simulations. Initial studies with autoencoders show promise in detecting unmodeled transients, termed ``anomalies'' by comparing reconstruction errors with predefined thresholds~\cite{Morawski:2021kxv, eric_moreno_2021_5772814}.

To date, machine learning (ML) integration into gravitational-wave detection pipelines has been limited but promising~\cite{verma2024detectiongravitationalwavesignals, PhysRevD.103.102003, Krastev_2020, PhysRevD.97.044039, PhysRevD.108.024022}. However, exploring the application of appropriate AI-based techniques for scientific discovery is a new direction that the scientific community is now turning towards. One such current ML-based algorithm in the LVK pipeline, MLy, has demonstrated utility in improving detection rates and reducing false positives~\cite{skliris2024realtimedetectionunmodelledgravitationalwave}. Despite its success, there remains a significant opportunity to enhance the generality and adaptability of anomaly detection in gravitational wave data.

In response to these challenges, this paper introduces the first NN-based analysis of O3 data in a search for unmodeled transients, using the novel semi-supervised machine learning method known as Gravitational Wave Anomalous Knowledge (\texttt{GWAK}). The \texttt{GWAK} method was first introduced by the authors of this paper in~\cite{Raikman:2023ktu}, leveraging the strengths of semi-supervised learning to construct a low-dimensional embedded space that captures the unique physical signatures of GW signals. This innovative approach not only facilitates the detection of known types of signals such as CBCs and detector glitches but also enhances the sensitivity to several plausible sources of short-duration GW transients (``bursts'') that have not yet been observed, such as core-collapse supernovae (CCSNe),  neutron star excitations, non-linear memory effects, or cosmic string cusps and kinks \cite{targeted_SN_O1-2, O2magnetarbursts, S6_NS, o2_mem,  O3cosmicstring}.  
Additional source populations could exist that are yet to be predicted. For these reasons, GW burst searches capable of detecting a wide range of signal waveforms provide a unique discovery opportunity.

Building on previous work, we train multiple autoencoders on different signals, including real background data, to improve performance. Our method differs from our prior work by utilizing real background data instead of simulated noise. The dataset used to train the \texttt{GWAK} autoencoder models in this study is based on the O3 background and simulated potential signals injected into it. See~\cite{Raikman:2023ktu} for greater details.

This paper presents the results of a generic all-sky search designed to detect a broad range of short-duration gravitational-wave bursts with diverse morphologies during O3. While this search is also sensitive to some CBC events \cite{allskyo2}, these are not the primary focus of the analysis, and detailed discussions of CBC detections in O3 can be found in \cite{LIGOScientific:2020ibl,O3IMBH}. After excluding CBC candidates, the search yields no statistically significant detections of other GW bursts.

This paper is structured to first cover the details of the analyzed dataset, then to outline the methodology behind the \texttt{GWAK} approach, followed by a detailed analysis of its application to O3. The ensuing sections will present experimental results, discuss the implications of these findings for gravitational wave astronomy, and suggest directions for future research.

\section{\label{sec:dataset} O3: The Third observing run}   
\subsection{Dataset}
The O3 data set extends from April 1, 2019, to March 27, 2020. We analyze the data collected by the Hanford (H1) and Livingston (L1) interferometers. 
The amount of data analyzed is reduced by requiring coincidence between two detectors and removing periods of critical data quality issues, as described further below. This results in a total of 203.3 analyzed days, which exceeds the 198.3 days analyzed in other O3 papers, such as~\cite{Abbott_2021}. Of the additional 5 days, 2.42 days come from the inclusion of Category 2 periods, while the remaining difference arises from algorithmic constraints and data segment limitations in~\cite{Abbott_2021}, such as the minimum segment length required by the algorithms.
The O3 GW strain data used in this paper is part of the O3 Data Release through the Gravitational Wave Open Science Center~\cite{Abbott_2023}.

\subsection{Data quality}
\label{ssec:DQ}
The LIGO and Virgo detectors are subject to various sources of terrestrial noise that can interfere with the detection of GWs~\cite{O3performance,Abbott_2016}. To monitor and mitigate these noise sources, the interferometers employ a large number of auxiliary channels that measure either environmental factors~\cite{envnoise,PEM} or internal interferometer behavior. Data quality (DQ) checks categorize noise contamination into two main classes:

\begin{itemize}
    \item \textbf{Category 1 (CAT1)}: Indicates critical issues where a key detector component or configuration is not functioning nominally. These segments are universally excluded from the analysis.
    \item \textbf{Category 2 (CAT2)}: Indicates times of known physical coupling to the gravitational wave channel, such as high seismic activity or other measurable disturbances.
\end{itemize}

Correlation with auxiliary channels is often used to identify CAT2 veto segments, marking times when noise transients are directly linked to these disturbances. While some analyses discard CAT2 segments entirely, we include these periods in our study to maximize the available observing time. This differs from the conventional approach, where discarding (vetoing) CAT2 times typically reduces the impact of noise transients but sacrifices additional data~\cite{dqmitigation}.

\section{\label{sec:methodology}Search algorithm}
\subsection{\label{sec:algorithm}\texttt{GWAK} Algorithm Overview}
The data analysis uses \texttt{GWAK}, a semi-supervised machine-learning framework to detect unmodeled gravitational-wave transients. \texttt{GWAK} leverages recurrent autoencoders to construct a low-dimensional embedded space where distinct features of gravitational wave signals, background noise, and glitches are represented. Five separate autoencoders (AEs) are trained on distinct data classes: background noise, glitches, and three signal types—Binary Black Holes and low- and high-frequency Sine-Gaussian (SG) waveforms. The encoded outputs form the axes of the \texttt{GWAK} space, where coherence features between input and reconstructed data are used to classify events. This semi-supervised approach incorporates signal priors while remaining general enough to detect previously unseen phenomena.

The \texttt{GWAK} algorithm evaluates new data segments by calculating the reconstruction loss from the five autoencoders and a frequency-domain correlation statistic. These features are combined using a linear classifier trained to distinguish signal-like events from background noise. By maintaining sensitivity across a wide range of signal morphologies, the method achieves robust performance in identifying CBCs and potential unmodeled sources, such as core-collapse supernovae and white-noise bursts. While the published~\cite{Raikman:2023ktu} paper focuses on the detailed description of the \texttt{GWAK} method, including its semi-supervised nature and the construction of a low-dimensional embedded space, this manuscript is dedicated to the analysis of real detector data.

\subsection{Training Dataset}

The dataset used in this study was collected by H1 and L1~\cite{TheLIGOScientific:2014jea} during O3. We specifically used publicly available data between GPS times of 1238166018--1253977218 (O3a) and GPS 125665561--1269363618 (O3b). 
Next, the time-series data were downsampled from 16384\,Hz to 4096\,Hz, and processed to remove and create a separate dataset of glitches using the excess power identification algorithm Omicron~\cite{Robinet:2020lbf}, using a Q$_{min} = 3.3166$, Q$_{max} = 108$, and f$_{min} = 32$ for the algorithm. The excised glitches were assigned to a dedicated AE class, while the remaining background data served as the background for injections into the three signal classes and was also used to train the final background AE. The BBH and SG sample is generated with the parameters and priors~\cite{alex_nitz_2020_3993665}, and as shown in Table~\ref{table:priors}. 

\begin{table}[htb]
\centering
\begin{tabular}{|l l l c l|}
\hline
\textbf{} & \textbf{Parameter} & \textbf{Prior} & \textbf{Limits} & \textbf{Units} \\ \hline
\multicolumn{5}{|c|}{\textbf{BBH}} \\ \hline
& $m_1$              & -       & $(5, 100)$           & $M_{\odot}$ \\
& $m_2$              & -       & $(5, 100)$           & $M_{\odot}$ \\
& Mass ratio $q$     & Uniform & $(0.125, 1)$         & - \\
& Chirp mass $M_c$   & Uniform & $(25, 100)$          & $M_{\odot}$ \\
& Tilts $\theta_{1,2}$ & Sine    & $(0, \pi)$           & rad. \\
& Phase $\phi$       & Uniform & $(0, 2\pi)$          & rad. \\
& Right Ascension    & Uniform & $(0, 2\pi)$          & rad. \\
& Declination $\delta$ & Cosine  & $(-\pi/2, \pi/2)$    & rad. \\ \hline
\multicolumn{5}{|c|}{\textbf{Sine-Gaussian}} \\ \hline
& $Q$                & Uniform & $(25, 75)$           & - \\
& Frequency          & Uniform & $(64, 512)$ and $(512, 1024)$ & Hz \\
& Phase $\phi$       & Uniform & $(0, 2\pi)$          & rad. \\
& Right Ascension    & Uniform & $(0, 2\pi)$          & rad. \\
& Declination $\delta$ & Cosine  & $(-\pi/2, \pi/2)$    & rad. \\
& Eccentricity       & Uniform & $(0, 0.01)$          & - \\
& $\Psi$             & Uniform & $(0, 2\pi)$          & rad. \\ \hline
\end{tabular}
\caption{Sampling parameters and priors for BBH (top) and SG (bottom) injections for \texttt{GWAK's} signal class AEs.}
\label{table:priors}
\end{table}

\subsection{Computational limitations}
The original \texttt{GWAK} method introduced in~\cite{Raikman:2023ktu}, also used the Pearson correlation coefficient as one of the anomaly axes.
However, computing the Pearson correlation between the two detectors is computationally expensive, as it requires evaluating 80 possible configurations at a sampling rate of 4096\,Hz. Each configuration corresponds to shifting the time axis of one detector within the range of [$-10$\,ms, $10$\,ms], reflecting the maximum possible light travel time between the detectors. To address this computational challenge, we utilized a frequency-domain correlation coefficient, defined as the dot product of the Fourier-transformed signals from each detector. This approach encodes equivalent information to the time-domain statistics while significantly reducing computational overhead. Since the Fourier transform is already performed for other parts of the analysis, this method adds negligible additional computational cost. Moreover, in the frequency domain, relative time shifts between the signals are inherently accounted for, eliminating the need for explicit time-axis sliding.

False alarm rates (FARs) are a standard metric used to quantify the statistical significance of detections across all GW search algorithms, including the \texttt{GWAK} algorithm. Achieving a significance level exceeding 3$\sigma$ at an FAR of 1 event per year requires evaluating the \texttt{GWAK} algorithm over approximately 10,000 years of timeslide data, posing significant computational challenges. By excluding the Pearson correlation from the final metric computation, this evaluation was completed in 4,500 GPU hours for the 10,000 years of timeslide data.

\subsection{\label{sec:detection_stat} Heuristic Model and Features}

In addition to the \texttt{GWAK} score introduced in the methods paper, we implement a heuristic model to address high false alarm rates observed during tests on 10,000 years of background data. Many significant false detections were associated with coincident glitches -- glitches appearing in both detectors with similar features within a $\sim 50$ ms window. Since the \texttt{GWAK} input segments are 50\,ms long, distinguishing such glitches from astrophysical signals required information outside this time window. Additionally, some false alarms arose from single-detector glitches misclassified by one of the signal autoencoders, as the training set primarily consisted of injections into both detectors and did not explicitly address this case.

\begin{figure}[htb]
\centering
\includegraphics[width=0.5\textwidth]{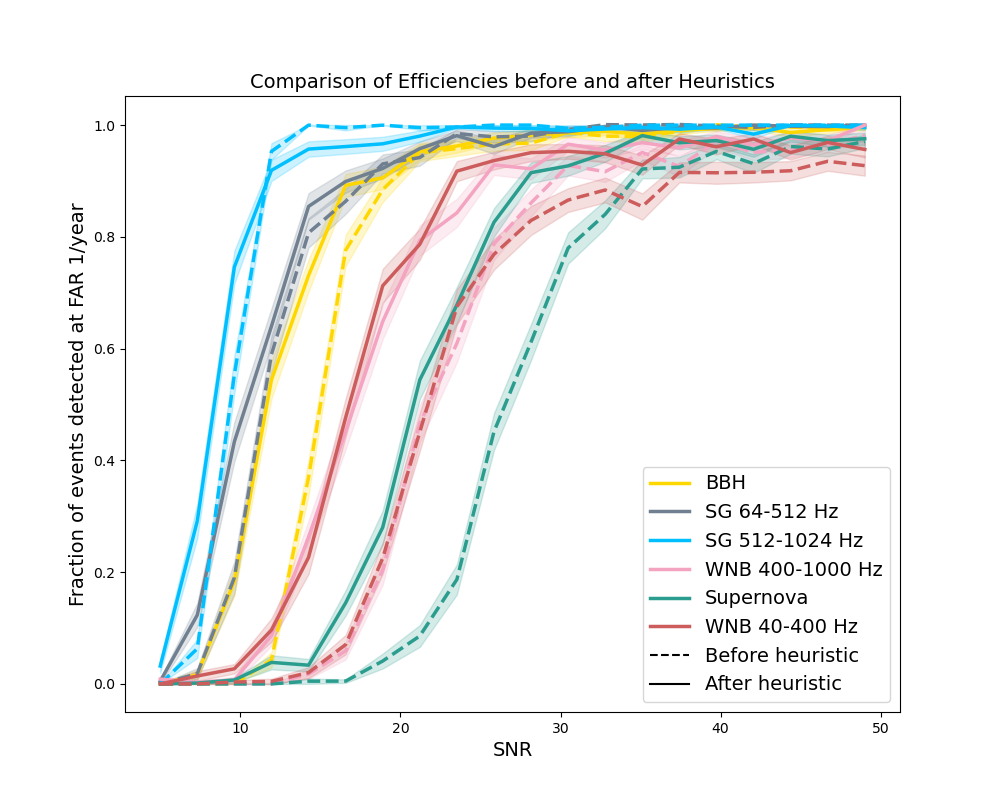} 
\caption{
The effect of applying the heuristic model on the signal efficiency at a false alarm rate of 1/year as a function of signal-to-noise ratio~(SNR). Solid lines represent efficiencies after applying the heuristic model, while dashed lines show efficiencies before applying the heuristic model. Each signal category is represented by a specific color: Binary Black Holes, BBH (yellow), low-frequency sine-Gaussian, SG 64--512 Hz (gray), high-frequency sine-Gaussian, SG 512--1024 Hz (blue), high-frequency white-noise bursts, WNB 400--1000 Hz (pink), Supernova (green), and low-frequency white-noise bursts, WNB 40--400 Hz (red).  Shaded regions correspond to statistical uncertainties. The heuristic model improves the separation of background noise from potential signals, leading to better detection performance.}
\label{fig:heuristics}
\end{figure}

To mitigate these issues, we developed a heuristic model trained as a binary classifier using engineered features. The model outputs a score between 0 and 1, where 0 represents a glitch and 1 represents a true signal. This score serves as a weight for the final detection statistic, obtained by multiplying the original \texttt{GWAK} scalar value with the heuristic score. Since the \texttt{GWAK} score for significant events is negative (with more negative values indicating stronger signals), a heuristic score closer to zero reduces the event’s significance, effectively down-weighting glitch-like events.

We used a set of six engineered features for the heuristic model. The first three were used to combat the problem with longer, coincident glitches that had similar morphology within a short, $\sim 50$\,ms window, but dissimilar morphology outside this $50$\,ms window. The first feature was the maximally time-shifted pearson correlation over a 1\,s window surrounding the event. Even if the coincident glitches overlapped for a short while, the dissimilar behavior in the rest of the 1\,s window would lead to a low pearson correlation. The next two features were a measure of the signal strength in either detector, taken as the log of the integral of the whitened PSD of the one-second surrounding window. The last three features quantified asymmetry in the scores given by the signal autoencoders. The formula to compute these scores is presented below.
\begin{equation}
    F_{k} = \max \left( \left|   \frac{G_{k, 0} + \Delta}{G_{k, 1} + \Delta} \right|, \left|\frac{G_{k, 1} + \Delta}{G_{k, 0} + \Delta}\right| \right)
\end{equation}

Here, $k$ corresponds to each signal autoencoder (BBH, SG low frequency, SG high frequency), $G_{k, 0}, G_{k, 1}$ correspond to the autoencoder score on the Hanford and Livingston detectors respectively, and $\Delta$ corresponds to a hyperparameter to avoid large values with small autoencoder scores from either detector. We used $\Delta=2$. The heuristic model had four neurons, with the latter three having shared weights. The first neuron would compute a linear combination of those first three features with a bias term, and perform a sigmoid activation. The next three neurons would each be given the asymmetry score for each autoencoder, perform a linear operation with bias, and apply a sigmoid activation. Finally, all these outputs from the sigmoid layers would be multiplied to give a final value. Intuitively, this architecture represents a test being performed for each heuristic value, and the output of the sigmoids being multiplied represents the fact that all tests must pass to be scored as an astrophysical signal. To train this model, we used a randomly sampled subset of the timeslides analyzed for the false alarm rate calculations, consisting of $\sim$ 1000 years of background events as our negative examples, and a set of injected astrophysical signals as our positive examples. We used a scaled sigmoid for the loss function, given below. The reason for this choice was to enable the model to make strong distinctions on signal or background events, as the same function would be used to re-weight the original \texttt{GWAK} score as described above. 
\begin{equation}
    S_L(x) = 1- \frac{1}{1+\exp(-L (x-0.5))}
\end{equation}
Where $x$ is the output of the heuristic model, L is the scaling factor, and $S(x)$ is the final value used to re-weight the score. In practice, we used $L=40$.

The signal efficiencies before and after the pre-trained heuristic model applied are shown in Fig.~\ref{fig:heuristics}. The background significance is drastically reduced, allowing for reasonable FAR estimates.

\subsection{Sensitivity to Generic Signal Morphologies}
To facilitate comparisons with other pipelines designed to search for burst signals, a set of \emph{ad hoc} waveforms, spanning diverse morphologies, is used to estimate the sensitivity to generic signals. The waveform families employed in this analysis include sine-Gaussian wavelets and band-limited white-noise bursts (WNB). SG signals are characterized by their central frequency $f_0$ and quality factor $Q$, which determine their duration. WNB signals are specified by their lower frequency bound $f_{low}$, bandwidth $\Delta f$, and duration $\tau$.

These \emph{ad hoc} signals are injected into the detector network across a range of amplitudes, quantified by the root-mean-squared strain amplitude ($h_\mathrm{rss}$), expressed as:
\[
h_\mathrm{rss} = \sqrt{\int_{-\infty}^{\infty} \left(h_+^2(t) + h_\times^2(t)\right) dt},
\]
where $h_+$ and $h_\times$ represent the signal polarizations in the source frame.

\begin{table}[htb]
\centering
\begin{tabular}{|c|c|}
\hline
\textbf{Morphology} & \textbf{GWAK} \\ \hline
\multicolumn{2}{|c|}{\textbf{sine-Gaussian wavelets}} \\ \hline
$f_0 = 70~\mathrm{Hz},~Q=3$ & 0.89 \\ 
$f_0 = 70~\mathrm{Hz},~Q=100$ & 1.83 \\ 
$f_0 = 235~\mathrm{Hz},~Q=100$ & 0.89 \\ 
$f_0 = 554~\mathrm{Hz},~Q=9$ & 0.89 \\ 
$f_0 = 849~\mathrm{Hz},~Q=3$ & 1.83 \\ 
$f_0 = 1304~\mathrm{Hz},~Q=9$ & 0.55 \\ 
$f_0 = 1615~\mathrm{Hz},~Q=100$ & 0.55 \\ 
$f_0 = 2000~\mathrm{Hz},~Q=3$ & 0.55 \\ \hline
\multicolumn{2}{|c|}{\textbf{White-noise bursts}} \\ \hline
$f_\mathrm{low} = 100~\mathrm{Hz},~\Delta f = 100~\mathrm{Hz},~\tau = 0.1~\mathrm{s}$ & 0.7 \\ 
$f_\mathrm{low} = 250~\mathrm{Hz},~\Delta f = 100~\mathrm{Hz},~\tau = 0.1~\mathrm{s}$ & 0.89 \\ \hline
\end{tabular}
\caption{The $h_\mathrm{rss}$ values, in units of $10^{-22}~\mathrm{Hz}^{-1/2}$, at which 50\% detection efficiency is achieved at a FAR of 1 in 100 years for the \texttt{GWAK} algorithm for various injected signal morphologies.}
\label{tab:hrss}
\end{table}

The \texttt{GWAK} search method is run in these signals to test the pipelines' capability to recover a wide range of transient signals. These signals include both a set of \emph{ad hoc} waveforms and astrophysically motivated waveforms originating from CCSNe and neutron star \emph{f}-modes. The $h_\mathrm{rss}$ values at which 50\% of signals are detected with $iFAR \geq 100$ years are provided in Table~\ref{tab:hrss}.

\section{\label{sec:results}Results of the GWTC-3 analysis}

\begin{figure}[htb]
\centering
\includegraphics[width=0.48\textwidth]{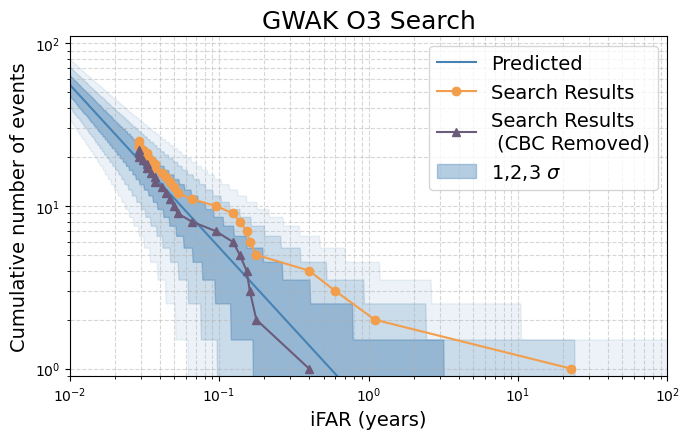}
\caption{
Cumulative number of events versus False Alarm rate (FAR) found by \texttt{GWAK} analysis using all O3 data. Circular points show results for all data and triangular points show after times around all known CBC sources have been excised. The solid line shows the expected mean value of the background, given the analyzed time. The shaded regions show the 1, 2, and 3 $\sigma$ Poisson uncertainty regions. The plot affirms that all \texttt{GWAK} detections during O3 are consistent with statistical fluctuations and CBC candidates.}
\label{fig:ifar}
\end{figure}

\begin{figure*}[htb]
\centering
\includegraphics[width=0.45\linewidth]{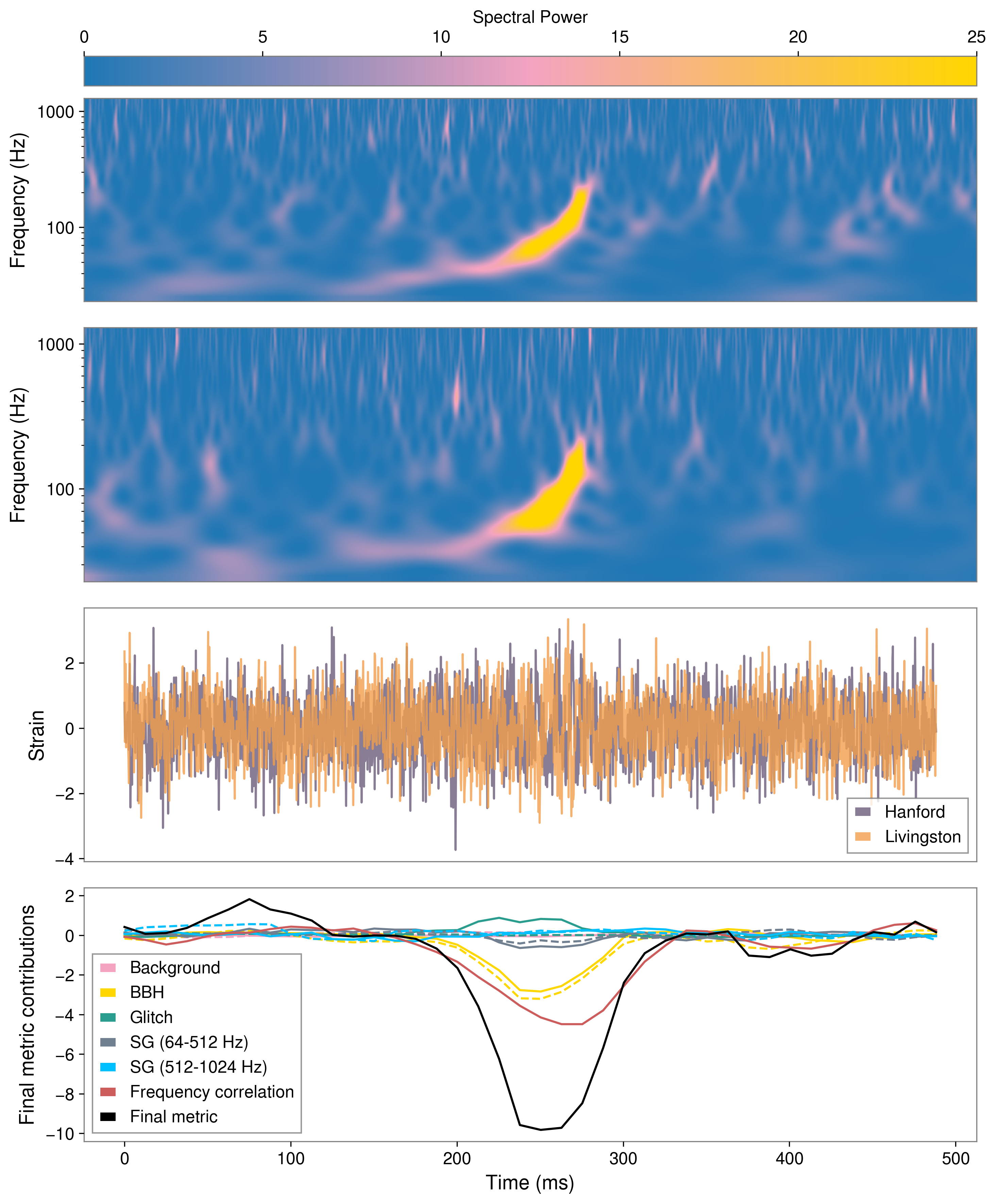} 
\includegraphics[width=0.45\linewidth]{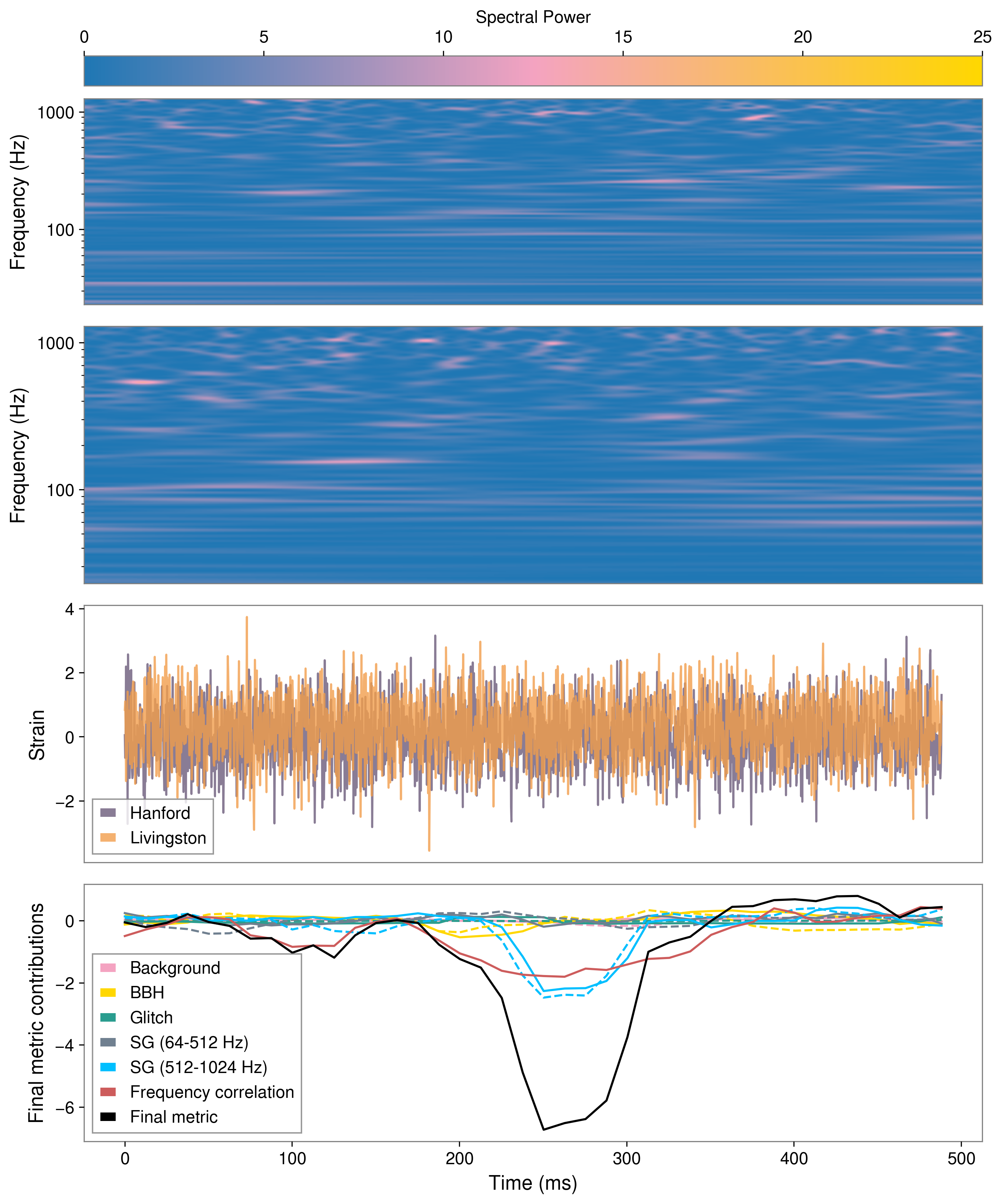} 
\caption{Example of a real BBH merger event~(left) and the loudest non-BBH detection~(right) identified by the \texttt{GWAK} algorithm in the O3 dataset. The top two panels display the spectrograms of the strain data from H1 and L1, highlighting the characteristic chirp signal of the event across the frequency range over time. The third panel shows the corresponding strain time series for both detectors, illustrating the amplitude evolution of the gravitational wave signal. The bottom panel provides the contributions of different \texttt{GWAK} features (e.g., background, glitch, BBH, sine-Gaussian signals, and frequency-domain correlation) to the final metric. The pronounced dip in the final metric (black curve) corresponds to the detection of the event. As expected, the largest contribution to the final metric for the BBH event~(left) originates from the BBH autoencoders, reflecting their design to reconstruct signals with morphology matching binary black hole mergers.}
\label{fig:bbh}
\end{figure*}

The analysis results for the \texttt{GWAK} are shown in Fig.~\ref{fig:ifar}.   
The loudest candidate event excluding known CBCs \cite{LIGOScientific:2020ibl} occurred at UTC May 05, 2019 15:10:38.
This candidate is shown in Fig.~\ref{fig:bbh}~(right) has an iFAR of approximately $1$~month. 
Though none of these meet an iFAR threshold indicating a potential detection, e.g. 1 per 100 years, an investigation into this loudest remaining candidate was conducted, concluding that this is a detector glitch. This cutoff was chosen to be consistent with ~\cite{Abbott_2021}.

\begin{figure*}[htb]
\centering
\includegraphics[width=0.8\textwidth]{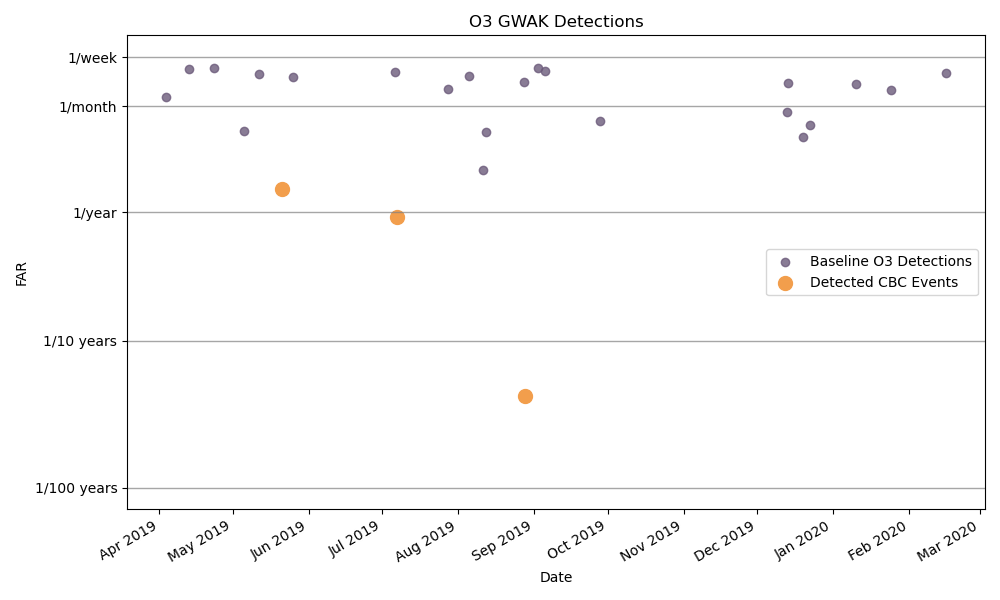}
\caption{
The distribution of events detected by the \texttt{GWAK} algorithm during the O3 of LIGO-Virgo-KAGRA, spanning from April 9, 2019, to March 21, 2020. The vertical axis represents the false alarm rate (FAR), indicating the significance of each detection, with higher FAR values (e.g., $>$\,1/week) corresponding to more significant events. The detections are categorized as baseline O3 detections (purple) and confirmed CBC events (orange). }
\label{fig:website}
\end{figure*}

Fig.~\ref{fig:website} illustrates the detection results from \texttt{GWAK} applied to O3. The plot displays the false alarm rates of identified events over time, categorized into baseline O3 detections and confirmed CBC events. This result underscores the effectiveness of \texttt{GWAK} in identifying signals of varying significance and its potential to complement traditional search pipelines in uncovering GW candidates.

\texttt{GWAK} identified multiple BBH signals with a high significance, the loudest one is shown in Fig.~\ref{fig:bbh}~(left). That event specifically was found and released by the LVK collaboration~\url{https://gwosc.org/eventapi/html/GWTC-2.1-confident/GW190828_063405/v2/}. It is important to notice that, as expected from the algorithm design, the BBH detection is mostly due to the BBH autoencoder ``firing'' and making the largest contribution to the total detection metric.

\begin{figure}[htb]
\centering
\includegraphics[width=0.48\textwidth]{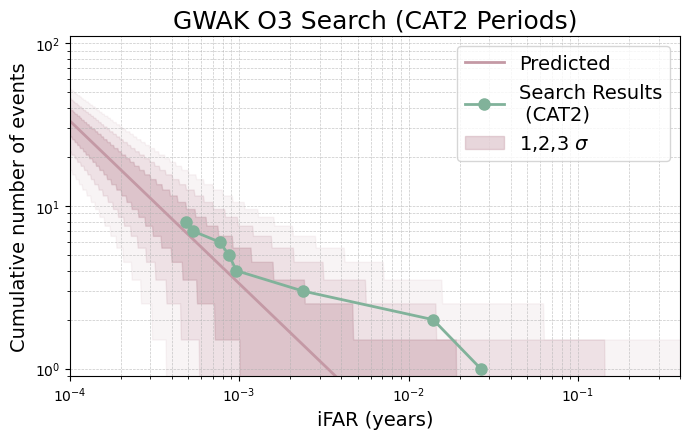}
\caption{
Cumulative number of events versus False Alarm rate (FAR) found by \texttt{GWAK} analysis using all CAT2 O3 data. Circular points show results for all CAT2 data. The solid line shows the expected mean value of the background, given the analyzed time. The shaded regions show the 1, 2, and 3 $\sigma$ Poisson uncertainty regions. The plot affirms that all \texttt{GWAK} detections during CAT2 periods are consistent with statistical fluctuations.}
\label{fig:ifar_cat2}
\end{figure}

\subsection{Analysis of a high-glitch (CAT2) periods}
\begin{figure*}[htb]
\includegraphics[width=0.45\linewidth]{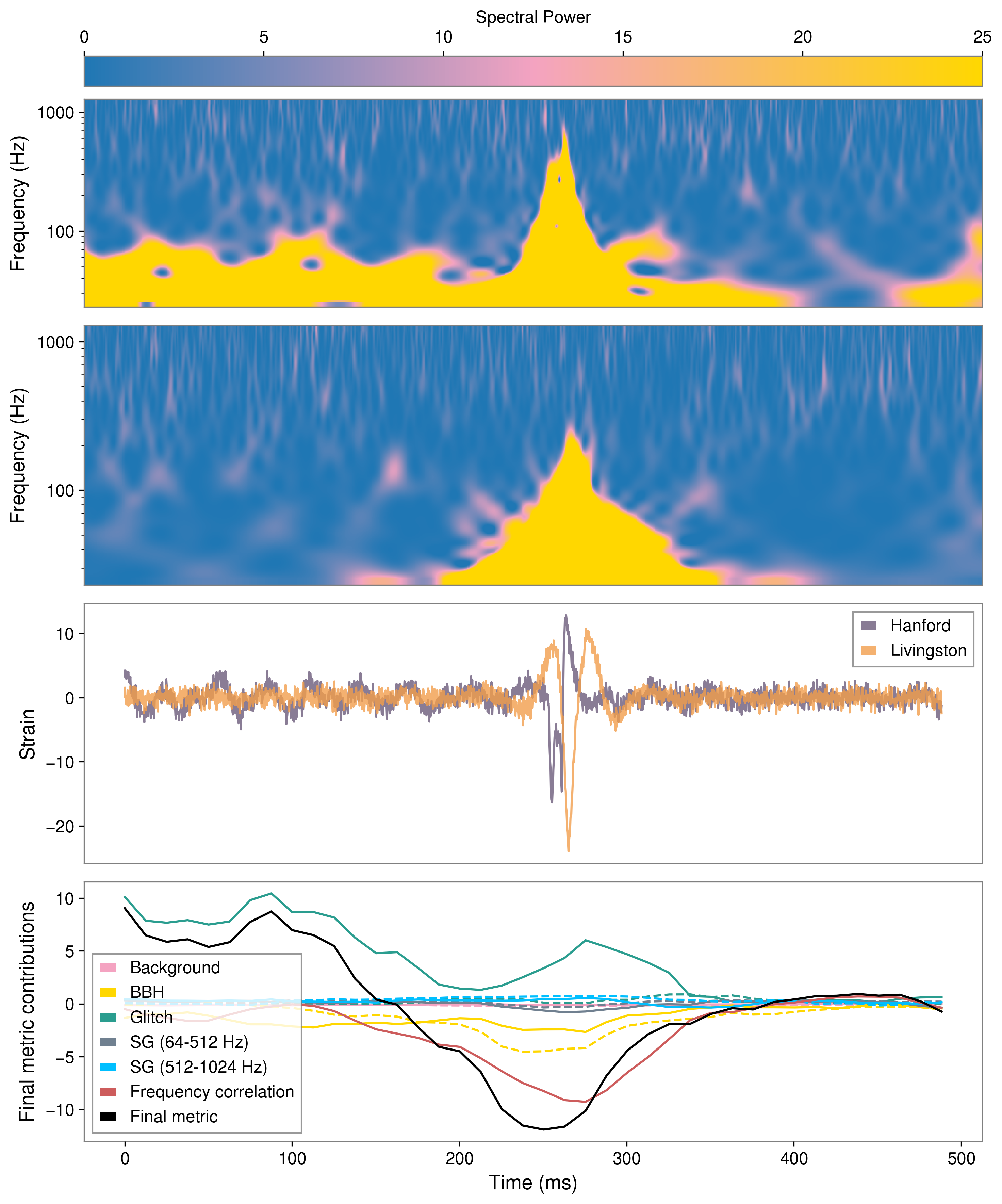} 
\includegraphics[width=0.45\linewidth]{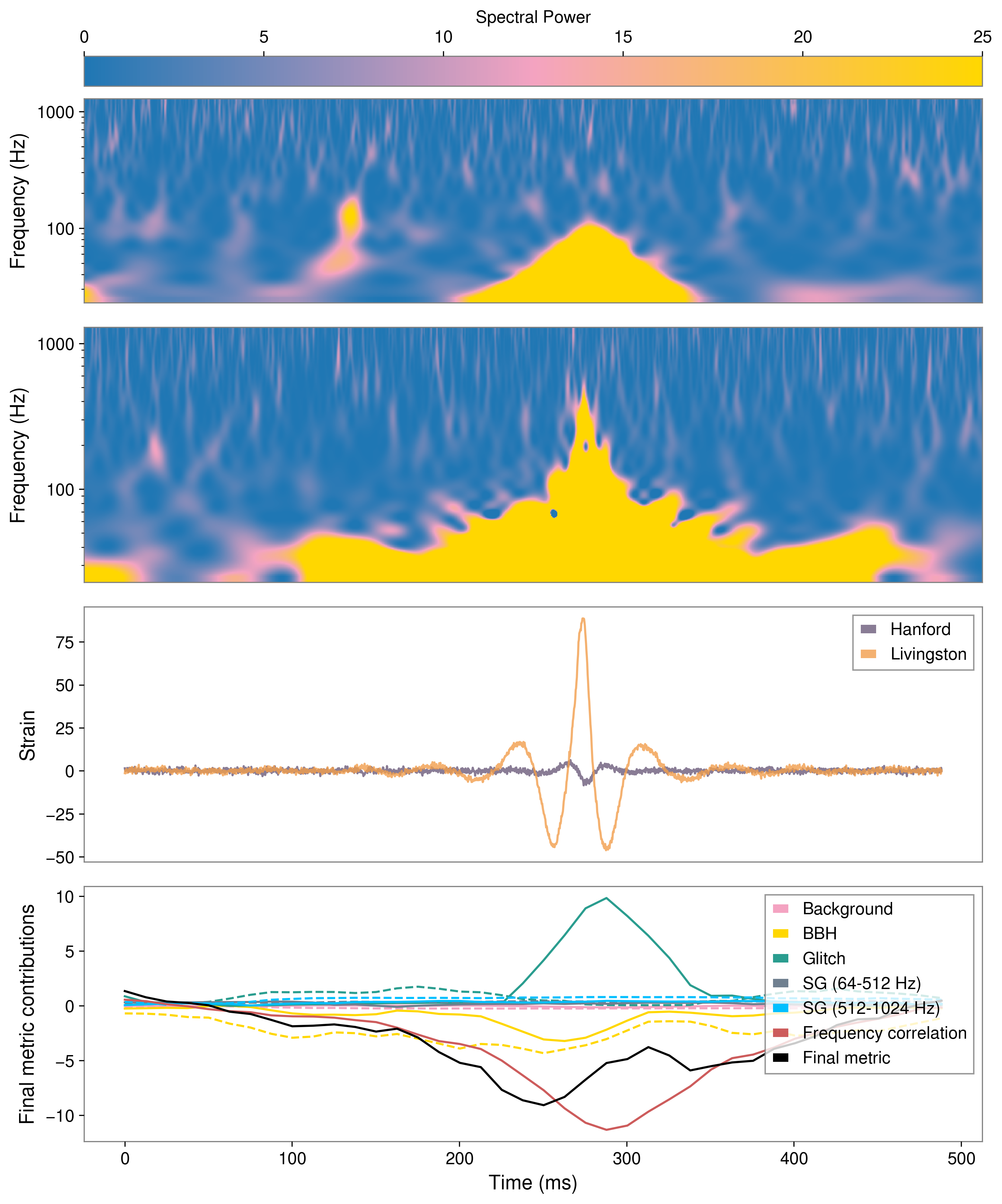} 
\caption{Example of the loudest~(left) and the second loudest~(right) anomalies found by the \texttt{GWAK} algorithm in the CAT2 dataset. The top two panels display the spectrograms of the strain data from H1 and L1, showcasing a transient signal with a distinctive high-frequency peak and rapid evolution over time. The third panel shows the strain time series for both detectors, highlighting the significant amplitude and temporal structure of the signal. The bottom panel provides the contributions of various \texttt{GWAK} features (e.g., background, glitch, BBH, sine-Gaussian signals, and frequency-domain correlation) to the final metric.}
\label{fig:anomaly_cat2}
\end{figure*}

Together with the main, analysis-ready dataset, we analyze the periods marked as having a high-confidence veto. While typical detection pipelines are designed to skip such datasets, including those flagged as CAT2 events due to the potential presence of loud glitches, environmental correlations, or engineered injections, our analysis with \texttt{GWAK} intentionally includes them. This approach is not about disregarding the quality of the data but rather about rigorously testing \texttt{GWAK}'s robustness and exploring its ability to identify genuine, anomalous signals even under more challenging conditions. The cumulative number of events and the loudest and second loudest anomalies are shown in Figs.~\ref{fig:ifar_cat2} and ~\ref{fig:anomaly_cat2} respectively.
As expected, background timeslides derived from 
 non-CAT2 data indicate these events as having very low FARs, while timeslides including CAT2 data lead to a revised local significance level of a few weeks for the anomalies.

\section{\label{sec:conclusion}Conclusion}

In this study, we presented the application of the Gravitational Wave Anomalous Knowledge (\texttt{GWAK}) method, a novel neural network-based algorithm, to the third observing run (O3) data from the LIGO-Virgo-KAGRA (LVK) collaboration. Using a semi-supervised machine-learning approach, \texttt{GWAK} proved effective in detecting unmodeled gravitational-wave transients with a wide range of shapes. Our results show that the algorithm can enhance traditional pipelines and tackle challenges in identifying unmodeled signals.

The \texttt{GWAK} analysis successfully identified multiple events consistent with compact binary coalescences (CBCs) reported by existing LVK pipelines. Notably, the algorithm efficiently handled data from high-glitch periods (CAT2), which are traditionally excluded from gravitational-wave searches, showcasing its resilience in challenging conditions. While no new significant astrophysical detections were made beyond the known CBC events, the method’s sensitivity to potential unmodeled signals, including burst-like events, highlights its potential for future discoveries.

Looking ahead, the \texttt{GWAK} framework provides a strong foundation for further exploration in gravitational-wave data analysis. Future efforts could include integrating additional signal priors, optimizing computational performance, and expanding its applicability to the upcoming observing runs. Furthermore, the ability to identify anomalies in high-glitch datasets opens the door to studying previously overlooked transient phenomena.

\begin{acknowledgments} 
The authors acknowledge support from the National Science Foundation with grant numbers OAC-2117997 and CSSI-1931469.
This research was undertaken with the support of the LIGO computational clusters.
MWC and SM also acknowledge support from the National Science Foundation with grant number PHY-2308862. EM acknowledges support from the National Science Foundation with grant number GRFP-2141064. This material is based upon work supported by NSF's LIGO Laboratory which is a major facility fully funded by the National Science Foundation. 
\end{acknowledgments}

\clearpage

\bibliography{main}

\begin{thebibliography}{59}%
\makeatletter
\providecommand \@ifxundefined [1]{%
 \@ifx{#1\undefined}
}%
\providecommand \@ifnum [1]{%
 \ifnum #1\expandafter \@firstoftwo
 \else \expandafter \@secondoftwo
 \fi
}%
\providecommand \@ifx [1]{%
 \ifx #1\expandafter \@firstoftwo
 \else \expandafter \@secondoftwo
 \fi
}%
\providecommand \natexlab [1]{#1}%
\providecommand \enquote  [1]{``#1''}%
\providecommand \bibnamefont  [1]{#1}%
\providecommand \bibfnamefont [1]{#1}%
\providecommand \citenamefont [1]{#1}%
\providecommand \href@noop [0]{\@secondoftwo}%
\providecommand \href [0]{\begingroup \@sanitize@url \@href}%
\providecommand \@href[1]{\@@startlink{#1}\@@href}%
\providecommand \@@href[1]{\endgroup#1\@@endlink}%
\providecommand \@sanitize@url [0]{\catcode `\\12\catcode `\$12\catcode `\&12\catcode `\#12\catcode `\^12\catcode `\_12\catcode `\%12\relax}%
\providecommand \@@startlink[1]{}%
\providecommand \@@endlink[0]{}%
\providecommand \url  [0]{\begingroup\@sanitize@url \@url }%
\providecommand \@url [1]{\endgroup\@href {#1}{\urlprefix }}%
\providecommand \urlprefix  [0]{URL }%
\providecommand \Eprint [0]{\href }%
\providecommand \doibase [0]{https://doi.org/}%
\providecommand \selectlanguage [0]{\@gobble}%
\providecommand \bibinfo  [0]{\@secondoftwo}%
\providecommand \bibfield  [0]{\@secondoftwo}%
\providecommand \translation [1]{[#1]}%
\providecommand \BibitemOpen [0]{}%
\providecommand \bibitemStop [0]{}%
\providecommand \bibitemNoStop [0]{.\EOS\space}%
\providecommand \EOS [0]{\spacefactor3000\relax}%
\providecommand \BibitemShut  [1]{\csname bibitem#1\endcsname}%
\let\auto@bib@innerbib\@empty
\bibitem [{\citenamefont {Abbott}\ \emph {et~al.}(2016{\natexlab{a}})\citenamefont {Abbott} \emph {et~al.}}]{PhysRevLett.116.061102}%
  \BibitemOpen
  \bibfield  {author} {\bibinfo {author} {\bibfnamefont {B.~P.}\ \bibnamefont {Abbott}} \emph {et~al.} (\bibinfo {collaboration} {{LIGO} Scientific Collaboration and Virgo Collaboration}),\ }\bibfield  {title} {\bibinfo {title} {Observation of gravitational waves from a binary black hole merger},\ }\href {https://doi.org/10.1103/PhysRevLett.116.061102} {\bibfield  {journal} {\bibinfo  {journal} {Phys. Rev. Lett.}\ }\textbf {\bibinfo {volume} {116}},\ \bibinfo {pages} {061102} (\bibinfo {year} {2016}{\natexlab{a}})}\BibitemShut {NoStop}%
\bibitem [{\citenamefont {Collaboration}\ and\ \citenamefont {et~al.}(2015)}]{Aasi_2015}%
  \BibitemOpen
  \bibfield  {author} {\bibinfo {author} {\bibfnamefont {T.~L.~S.}\ \bibnamefont {Collaboration}}\ and\ \bibinfo {author} {\bibfnamefont {J.~A.}\ \bibnamefont {et~al.}},\ }\bibfield  {title} {\bibinfo {title} {Advanced {LIGO}},\ }\href {https://doi.org/10.1088/0264-9381/32/7/074001} {\bibfield  {journal} {\bibinfo  {journal} {Classical and Quantum Gravity}\ }\textbf {\bibinfo {volume} {32}},\ \bibinfo {pages} {074001} (\bibinfo {year} {2015})}\BibitemShut {NoStop}%
\bibitem [{\citenamefont {et~al.}(2014)}]{Acernese_2015}%
  \BibitemOpen
  \bibfield  {author} {\bibinfo {author} {\bibfnamefont {F.~A.}\ \bibnamefont {et~al.}},\ }\bibfield  {title} {\bibinfo {title} {Advanced virgo: a second-generation interferometric gravitational wave detector},\ }\href {https://doi.org/10.1088/0264-9381/32/2/024001} {\bibfield  {journal} {\bibinfo  {journal} {Classical and Quantum Gravity}\ }\textbf {\bibinfo {volume} {32}},\ \bibinfo {pages} {024001} (\bibinfo {year} {2014})}\BibitemShut {NoStop}%
\bibitem [{\citenamefont {Aso}\ \emph {et~al.}(2013)\citenamefont {Aso}, \citenamefont {Michimura}, \citenamefont {Somiya}, \citenamefont {Ando}, \citenamefont {Miyakawa}, \citenamefont {Sekiguchi}, \citenamefont {Tatsumi},\ and\ \citenamefont {Yamamoto}}]{Aso_2013}%
  \BibitemOpen
  \bibfield  {author} {\bibinfo {author} {\bibfnamefont {Y.}~\bibnamefont {Aso}}, \bibinfo {author} {\bibfnamefont {Y.}~\bibnamefont {Michimura}}, \bibinfo {author} {\bibfnamefont {K.}~\bibnamefont {Somiya}}, \bibinfo {author} {\bibfnamefont {M.}~\bibnamefont {Ando}}, \bibinfo {author} {\bibfnamefont {O.}~\bibnamefont {Miyakawa}}, \bibinfo {author} {\bibfnamefont {T.}~\bibnamefont {Sekiguchi}}, \bibinfo {author} {\bibfnamefont {D.}~\bibnamefont {Tatsumi}},\ and\ \bibinfo {author} {\bibfnamefont {H.}~\bibnamefont {Yamamoto}},\ }\bibfield  {title} {\bibinfo {title} {Interferometer design of the kagra gravitational wave detector},\ }\bibfield  {journal} {\bibinfo  {journal} {Physical Review D}\ }\textbf {\bibinfo {volume} {88}},\ \href {https://doi.org/10.1103/physrevd.88.043007} {10.1103/physrevd.88.043007} (\bibinfo {year} {2013})\BibitemShut {NoStop}%
\bibitem [{\citenamefont {Abbott}\ \emph {et~al.}(2021{\natexlab{a}})\citenamefont {Abbott} \emph {et~al.}}]{LIGOScientific:2020ibl}%
  \BibitemOpen
  \bibfield  {author} {\bibinfo {author} {\bibfnamefont {R.}~\bibnamefont {Abbott}} \emph {et~al.} (\bibinfo {collaboration} {LIGO Scientific, Virgo}),\ }\bibfield  {title} {\bibinfo {title} {{GWTC-2: Compact Binary Coalescences Observed by {LIGO} and Virgo During the First Half of the Third Observing Run}},\ }\href {https://doi.org/10.1103/PhysRevX.11.021053} {\bibfield  {journal} {\bibinfo  {journal} {Phys. Rev. X}\ }\textbf {\bibinfo {volume} {11}},\ \bibinfo {pages} {021053} (\bibinfo {year} {2021}{\natexlab{a}})},\ \Eprint {https://arxiv.org/abs/2010.14527} {arXiv:2010.14527 [gr-qc]} \BibitemShut {NoStop}%
\bibitem [{\citenamefont {Abbott}\ \emph {et~al.}(2024)\citenamefont {Abbott} \emph {et~al.}}]{LIGOScientific:2021usb}%
  \BibitemOpen
  \bibfield  {author} {\bibinfo {author} {\bibfnamefont {R.}~\bibnamefont {Abbott}} \emph {et~al.} (\bibinfo {collaboration} {LIGO Scientific, VIRGO}),\ }\bibfield  {title} {\bibinfo {title} {{GWTC-2.1: Deep extended catalog of compact binary coalescences observed by {LIGO} and Virgo during the first half of the third observing run}},\ }\href {https://doi.org/10.1103/PhysRevD.109.022001} {\bibfield  {journal} {\bibinfo  {journal} {Phys. Rev. D}\ }\textbf {\bibinfo {volume} {109}},\ \bibinfo {pages} {022001} (\bibinfo {year} {2024})},\ \Eprint {https://arxiv.org/abs/2108.01045} {arXiv:2108.01045 [gr-qc]} \BibitemShut {NoStop}%
\bibitem [{\citenamefont {Abbott}\ \emph {et~al.}(2019{\natexlab{a}})\citenamefont {Abbott}, \citenamefont {Abbott},\ and\ \citenamefont {Abbott}}]{PhysRevX.9.031040}%
  \BibitemOpen
  \bibfield  {author} {\bibinfo {author} {\bibfnamefont {B.~P.}\ \bibnamefont {Abbott}}, \bibinfo {author} {\bibfnamefont {R.}~\bibnamefont {Abbott}},\ and\ \bibinfo {author} {\bibfnamefont {T.~D. e.~a.}\ \bibnamefont {Abbott}} (\bibinfo {collaboration} {LIGO Scientific Collaboration and Virgo Collaboration}),\ }\bibfield  {title} {\bibinfo {title} {Gwtc-1: A gravitational-wave transient catalog of compact binary mergers observed by {LIGO} and virgo during the first and second observing runs},\ }\href {https://doi.org/10.1103/PhysRevX.9.031040} {\bibfield  {journal} {\bibinfo  {journal} {Phys. Rev. X}\ }\textbf {\bibinfo {volume} {9}},\ \bibinfo {pages} {031040} (\bibinfo {year} {2019}{\natexlab{a}})}\BibitemShut {NoStop}%
\bibitem [{\citenamefont {Helstrom}(1968)}]{helstrom1968}%
  \BibitemOpen
  \bibfield  {author} {\bibinfo {author} {\bibfnamefont {C.~W.}\ \bibnamefont {Helstrom}},\ }\href {https://search.ebscohost.com/login.aspx?direct=true&scope=site&db=nlebk&db=nlabk&AN=881839} {\emph {\bibinfo {title} {Statistical Theory of Signal Detection}}},\ \bibinfo {edition} {2nd}\ ed.\ (\bibinfo  {publisher} {Pergamon Press},\ \bibinfo {year} {1968})\BibitemShut {NoStop}%
\bibitem [{\citenamefont {Allen}\ \emph {et~al.}(2012)\citenamefont {Allen} \emph {et~al.}}]{Allen:2005fk}%
  \BibitemOpen
  \bibfield  {author} {\bibinfo {author} {\bibfnamefont {B.}~\bibnamefont {Allen}} \emph {et~al.},\ }\bibfield  {title} {\bibinfo {title} {{FINDCHIRP: An Algorithm for detection of gravitational waves from inspiraling compact binaries}},\ }\href {https://doi.org/10.1103/PhysRevD.85.122006} {\bibfield  {journal} {\bibinfo  {journal} {Phys. Rev. D}\ }\textbf {\bibinfo {volume} {85}},\ \bibinfo {pages} {122006} (\bibinfo {year} {2012})},\ \Eprint {https://arxiv.org/abs/gr-qc/0509116} {arXiv:gr-qc/0509116} \BibitemShut {NoStop}%
\bibitem [{\citenamefont {Klimenko}\ \emph {et~al.}(2008)\citenamefont {Klimenko}, \citenamefont {Yakushin}, \citenamefont {Mercer},\ and\ \citenamefont {Mitselmakher}}]{Klimenko:2008fu}%
  \BibitemOpen
  \bibfield  {author} {\bibinfo {author} {\bibfnamefont {S.}~\bibnamefont {Klimenko}}, \bibinfo {author} {\bibfnamefont {I.}~\bibnamefont {Yakushin}}, \bibinfo {author} {\bibfnamefont {A.}~\bibnamefont {Mercer}},\ and\ \bibinfo {author} {\bibfnamefont {G.}~\bibnamefont {Mitselmakher}},\ }\bibfield  {title} {\bibinfo {title} {{Coherent method for detection of gravitational wave bursts}},\ }\href {https://doi.org/10.1088/0264-9381/25/11/114029} {\bibfield  {journal} {\bibinfo  {journal} {Class. Quant. Grav.}\ }\textbf {\bibinfo {volume} {25}},\ \bibinfo {pages} {114029} (\bibinfo {year} {2008})},\ \Eprint {https://arxiv.org/abs/0802.3232} {arXiv:0802.3232 [gr-qc]} \BibitemShut {NoStop}%
\bibitem [{\citenamefont {Lynch}\ \emph {et~al.}(2017)\citenamefont {Lynch}, \citenamefont {Vitale}, \citenamefont {Essick}, \citenamefont {Katsavounidis},\ and\ \citenamefont {Robinet}}]{Lynch:2015yin}%
  \BibitemOpen
  \bibfield  {author} {\bibinfo {author} {\bibfnamefont {R.}~\bibnamefont {Lynch}}, \bibinfo {author} {\bibfnamefont {S.}~\bibnamefont {Vitale}}, \bibinfo {author} {\bibfnamefont {R.}~\bibnamefont {Essick}}, \bibinfo {author} {\bibfnamefont {E.}~\bibnamefont {Katsavounidis}},\ and\ \bibinfo {author} {\bibfnamefont {F.}~\bibnamefont {Robinet}},\ }\bibfield  {title} {\bibinfo {title} {{Information-theoretic approach to the gravitational-wave burst detection problem}},\ }\href {https://doi.org/10.1103/PhysRevD.95.104046} {\bibfield  {journal} {\bibinfo  {journal} {Phys. Rev. D}\ }\textbf {\bibinfo {volume} {95}},\ \bibinfo {pages} {104046} (\bibinfo {year} {2017})},\ \Eprint {https://arxiv.org/abs/1511.05955} {arXiv:1511.05955 [gr-qc]} \BibitemShut {NoStop}%
\bibitem [{\citenamefont {Macquet}\ \emph {et~al.}(2021)\citenamefont {Macquet}, \citenamefont {Bizouard}, \citenamefont {Christensen},\ and\ \citenamefont {Coughlin}}]{Macquet:2021ibe}%
  \BibitemOpen
  \bibfield  {author} {\bibinfo {author} {\bibfnamefont {A.}~\bibnamefont {Macquet}}, \bibinfo {author} {\bibfnamefont {M.-A.}\ \bibnamefont {Bizouard}}, \bibinfo {author} {\bibfnamefont {N.}~\bibnamefont {Christensen}},\ and\ \bibinfo {author} {\bibfnamefont {M.}~\bibnamefont {Coughlin}},\ }\bibfield  {title} {\bibinfo {title} {{Searches for long-duration gravitational wave transients in {LIGO} and Virgo data}},\ }in\ \href@noop {} {\emph {\bibinfo {booktitle} {{55th Rencontres de Moriond on Gravitation}}}}\ (\bibinfo {year} {2021})\ \Eprint {https://arxiv.org/abs/2108.10033} {arXiv:2108.10033 [gr-qc]} \BibitemShut {NoStop}%
\bibitem [{\citenamefont {Cornish}\ and\ \citenamefont {Littenberg}(2015)}]{Cornish:2014kda}%
  \BibitemOpen
  \bibfield  {author} {\bibinfo {author} {\bibfnamefont {N.~J.}\ \bibnamefont {Cornish}}\ and\ \bibinfo {author} {\bibfnamefont {T.~B.}\ \bibnamefont {Littenberg}},\ }\bibfield  {title} {\bibinfo {title} {{BayesWave: Bayesian Inference for Gravitational Wave Bursts and Instrument Glitches}},\ }\href {https://doi.org/10.1088/0264-9381/32/13/135012} {\bibfield  {journal} {\bibinfo  {journal} {Class. Quant. Grav.}\ }\textbf {\bibinfo {volume} {32}},\ \bibinfo {pages} {135012} (\bibinfo {year} {2015})},\ \Eprint {https://arxiv.org/abs/1410.3835} {arXiv:1410.3835 [gr-qc]} \BibitemShut {NoStop}%
\bibitem [{\citenamefont {The LIGO Scientific~Collaboration}\ and\ \citenamefont {the KAGRA~Collaboration}(2021)}]{Abbott_2021}%
  \BibitemOpen
  \bibfield  {author} {\bibinfo {author} {\bibfnamefont {t.~V.~C.}\ \bibnamefont {The LIGO Scientific~Collaboration}}\ and\ \bibinfo {author} {\bibnamefont {the KAGRA~Collaboration}},\ }\bibfield  {title} {\bibinfo {title} {All-sky search for short gravitational-wave bursts in the third advanced {LIGO} and advanced virgo run},\ }\bibfield  {journal} {\bibinfo  {journal} {Physical Review D}\ }\textbf {\bibinfo {volume} {104}},\ \href {https://doi.org/10.1103/physrevd.104.122004} {10.1103/physrevd.104.122004} (\bibinfo {year} {2021})\BibitemShut {NoStop}%
\bibitem [{\citenamefont {Abbott}\ \emph {et~al.}(2021{\natexlab{b}})\citenamefont {Abbott} \emph {et~al.}}]{KAGRA:2021bhs}%
  \BibitemOpen
  \bibfield  {author} {\bibinfo {author} {\bibfnamefont {R.}~\bibnamefont {Abbott}} \emph {et~al.} (\bibinfo {collaboration} {KAGRA, VIRGO, {LIGO} Scientific}),\ }\bibfield  {title} {\bibinfo {title} {{All-sky search for long-duration gravitational-wave bursts in the third Advanced {LIGO} and Advanced Virgo run}},\ }\href {https://doi.org/10.1103/PhysRevD.104.102001} {\bibfield  {journal} {\bibinfo  {journal} {Phys. Rev. D}\ }\textbf {\bibinfo {volume} {104}},\ \bibinfo {pages} {102001} (\bibinfo {year} {2021}{\natexlab{b}})},\ \Eprint {https://arxiv.org/abs/2107.13796} {arXiv:2107.13796 [gr-qc]} \BibitemShut {NoStop}%
\bibitem [{\citenamefont {Baker}\ \emph {et~al.}(2015)\citenamefont {Baker}, \citenamefont {Caudill}, \citenamefont {Hodge}, \citenamefont {Talukder}, \citenamefont {Capano},\ and\ \citenamefont {Cornish}}]{Baker:2014eba}%
  \BibitemOpen
  \bibfield  {author} {\bibinfo {author} {\bibfnamefont {P.~T.}\ \bibnamefont {Baker}}, \bibinfo {author} {\bibfnamefont {S.}~\bibnamefont {Caudill}}, \bibinfo {author} {\bibfnamefont {K.~A.}\ \bibnamefont {Hodge}}, \bibinfo {author} {\bibfnamefont {D.}~\bibnamefont {Talukder}}, \bibinfo {author} {\bibfnamefont {C.}~\bibnamefont {Capano}},\ and\ \bibinfo {author} {\bibfnamefont {N.~J.}\ \bibnamefont {Cornish}},\ }\bibfield  {title} {\bibinfo {title} {{Multivariate Classification with Random Forests for Gravitational Wave Searches of Black Hole Binary Coalescence}},\ }\href {https://doi.org/10.1103/PhysRevD.91.062004} {\bibfield  {journal} {\bibinfo  {journal} {Phys. Rev. D}\ }\textbf {\bibinfo {volume} {91}},\ \bibinfo {pages} {062004} (\bibinfo {year} {2015})},\ \Eprint {https://arxiv.org/abs/1412.6479} {arXiv:1412.6479 [gr-qc]} \BibitemShut {NoStop}%
\bibitem [{\citenamefont {George}\ and\ \citenamefont {Huerta}(2018{\natexlab{a}})}]{George:2016hay}%
  \BibitemOpen
  \bibfield  {author} {\bibinfo {author} {\bibfnamefont {D.}~\bibnamefont {George}}\ and\ \bibinfo {author} {\bibfnamefont {E.}~\bibnamefont {Huerta}},\ }\bibfield  {title} {\bibinfo {title} {{Deep Neural Networks to Enable Real-time Multimessenger Astrophysics}},\ }\href {https://doi.org/10.1103/PhysRevD.97.044039} {\bibfield  {journal} {\bibinfo  {journal} {Phys. Rev. D}\ }\textbf {\bibinfo {volume} {97}},\ \bibinfo {pages} {044039} (\bibinfo {year} {2018}{\natexlab{a}})},\ \Eprint {https://arxiv.org/abs/1701.00008} {arXiv:1701.00008 [astro-ph.IM]} \BibitemShut {NoStop}%
\bibitem [{\citenamefont {Kapadia}\ \emph {et~al.}(2017)\citenamefont {Kapadia}, \citenamefont {Dent},\ and\ \citenamefont {Dal~Canton}}]{Kapadia:2017fhb}%
  \BibitemOpen
  \bibfield  {author} {\bibinfo {author} {\bibfnamefont {S.~J.}\ \bibnamefont {Kapadia}}, \bibinfo {author} {\bibfnamefont {T.}~\bibnamefont {Dent}},\ and\ \bibinfo {author} {\bibfnamefont {T.}~\bibnamefont {Dal~Canton}},\ }\bibfield  {title} {\bibinfo {title} {{Classifier for gravitational-wave inspiral signals in nonideal single-detector data}},\ }\href {https://doi.org/10.1103/PhysRevD.96.104015} {\bibfield  {journal} {\bibinfo  {journal} {Phys. Rev. D}\ }\textbf {\bibinfo {volume} {96}},\ \bibinfo {pages} {104015} (\bibinfo {year} {2017})},\ \Eprint {https://arxiv.org/abs/1709.02421} {arXiv:1709.02421 [astro-ph.IM]} \BibitemShut {NoStop}%
\bibitem [{\citenamefont {George}\ and\ \citenamefont {Huerta}(2018{\natexlab{b}})}]{George:2017pmj}%
  \BibitemOpen
  \bibfield  {author} {\bibinfo {author} {\bibfnamefont {D.}~\bibnamefont {George}}\ and\ \bibinfo {author} {\bibfnamefont {E.}~\bibnamefont {Huerta}},\ }\bibfield  {title} {\bibinfo {title} {{Deep Learning for Real-time Gravitational Wave Detection and Parameter Estimation: Results with Advanced {LIGO} Data}},\ }\href {https://doi.org/10.1016/j.physletb.2017.12.053} {\bibfield  {journal} {\bibinfo  {journal} {Phys. Lett. B}\ }\textbf {\bibinfo {volume} {778}},\ \bibinfo {pages} {64} (\bibinfo {year} {2018}{\natexlab{b}})},\ \Eprint {https://arxiv.org/abs/1711.03121} {arXiv:1711.03121 [gr-qc]} \BibitemShut {NoStop}%
\bibitem [{\citenamefont {Gabbard}\ \emph {et~al.}(2018)\citenamefont {Gabbard} \emph {et~al.}}]{Gabbard:2017lja}%
  \BibitemOpen
  \bibfield  {author} {\bibinfo {author} {\bibfnamefont {H.}~\bibnamefont {Gabbard}} \emph {et~al.},\ }\bibfield  {title} {\bibinfo {title} {{Matching matched filtering with deep networks for gravitational-wave astronomy}},\ }\href {https://doi.org/10.1103/PhysRevLett.120.141103} {\bibfield  {journal} {\bibinfo  {journal} {Phys. Rev. Lett.}\ }\textbf {\bibinfo {volume} {120}},\ \bibinfo {pages} {141103} (\bibinfo {year} {2018})},\ \Eprint {https://arxiv.org/abs/1712.06041} {arXiv:1712.06041 [astro-ph.IM]} \BibitemShut {NoStop}%
\bibitem [{\citenamefont {Miller}\ \emph {et~al.}(2019)\citenamefont {Miller} \emph {et~al.}}]{Miller:2019jtp}%
  \BibitemOpen
  \bibfield  {author} {\bibinfo {author} {\bibfnamefont {A.~L.}\ \bibnamefont {Miller}} \emph {et~al.},\ }\bibfield  {title} {\bibinfo {title} {{How effective is machine learning to detect long transient gravitational waves from neutron stars in a real search?}},\ }\href {https://doi.org/10.1103/PhysRevD.100.062005} {\bibfield  {journal} {\bibinfo  {journal} {Phys. Rev. D}\ }\textbf {\bibinfo {volume} {100}},\ \bibinfo {pages} {062005} (\bibinfo {year} {2019})},\ \Eprint {https://arxiv.org/abs/1909.02262} {arXiv:1909.02262 [astro-ph.IM]} \BibitemShut {NoStop}%
\bibitem [{\citenamefont {Jadhav}\ \emph {et~al.}(2021)\citenamefont {Jadhav}, \citenamefont {Mukund}, \citenamefont {Gadre}, \citenamefont {Mitra},\ and\ \citenamefont {Abraham}}]{Jadhav:2020oyt}%
  \BibitemOpen
  \bibfield  {author} {\bibinfo {author} {\bibfnamefont {S.}~\bibnamefont {Jadhav}}, \bibinfo {author} {\bibfnamefont {N.}~\bibnamefont {Mukund}}, \bibinfo {author} {\bibfnamefont {B.}~\bibnamefont {Gadre}}, \bibinfo {author} {\bibfnamefont {S.}~\bibnamefont {Mitra}},\ and\ \bibinfo {author} {\bibfnamefont {S.}~\bibnamefont {Abraham}},\ }\bibfield  {title} {\bibinfo {title} {{Improving significance of binary black hole mergers in Advanced {LIGO} data using deep learning: Confirmation of GW151216}},\ }\href {https://doi.org/10.1103/PhysRevD.104.064051} {\bibfield  {journal} {\bibinfo  {journal} {Phys. Rev. D}\ }\textbf {\bibinfo {volume} {104}},\ \bibinfo {pages} {064051} (\bibinfo {year} {2021})},\ \Eprint {https://arxiv.org/abs/2010.08584} {arXiv:2010.08584 [gr-qc]} \BibitemShut {NoStop}%
\bibitem [{\citenamefont {Huerta}\ \emph {et~al.}(2021)\citenamefont {Huerta} \emph {et~al.}}]{Huerta:2020xyq}%
  \BibitemOpen
  \bibfield  {author} {\bibinfo {author} {\bibfnamefont {E.~A.}\ \bibnamefont {Huerta}} \emph {et~al.},\ }\bibfield  {title} {\bibinfo {title} {{Accelerated, scalable and reproducible AI-driven gravitational wave detection}},\ }\href {https://doi.org/10.1038/s41550-021-01405-0} {\bibfield  {journal} {\bibinfo  {journal} {Nature Astron.}\ }\textbf {\bibinfo {volume} {5}},\ \bibinfo {pages} {1062} (\bibinfo {year} {2021})},\ \Eprint {https://arxiv.org/abs/2012.08545} {arXiv:2012.08545 [gr-qc]} \BibitemShut {NoStop}%
\bibitem [{\citenamefont {Jiang}\ and\ \citenamefont {Luo}(2022)}]{9956104}%
  \BibitemOpen
  \bibfield  {author} {\bibinfo {author} {\bibfnamefont {L.}~\bibnamefont {Jiang}}\ and\ \bibinfo {author} {\bibfnamefont {Y.}~\bibnamefont {Luo}},\ }\bibfield  {title} {\bibinfo {title} {Convolutional transformer for fast and accurate gravitational wave detection},\ }in\ \href {https://doi.org/10.1109/ICPR56361.2022.9956104} {\emph {\bibinfo {booktitle} {2022 26th International Conference on Pattern Recognition (ICPR)}}}\ (\bibinfo {year} {2022})\ pp.\ \bibinfo {pages} {46--53}\BibitemShut {NoStop}%
\bibitem [{\citenamefont {{Beveridge, Damon and McLeod, Alistair and Wen, Linqing and Wicenec, Andreas}}(2023)}]{Beveridge:2023bxa}%
  \BibitemOpen
  \bibfield  {author} {\bibinfo {author} {\bibnamefont {{Beveridge, Damon and McLeod, Alistair and Wen, Linqing and Wicenec, Andreas}}},\ }\bibfield  {title} {\bibinfo {title} {{A Novel Deep Learning Approach to Detecting Binary Black Hole Mergers}},\ }\href@noop {} {\  (\bibinfo {year} {{2023}})},\ \Eprint {https://arxiv.org/abs/{2308.08429}} {{arXiv}:{2308.08429} [{gr-qc}]} \BibitemShut {NoStop}%
\bibitem [{\citenamefont {Marx}\ \emph {et~al.}(2024)\citenamefont {Marx} \emph {et~al.}}]{Marx:2024wjt}%
  \BibitemOpen
  \bibfield  {author} {\bibinfo {author} {\bibfnamefont {E.}~\bibnamefont {Marx}} \emph {et~al.},\ }\bibfield  {title} {\bibinfo {title} {{A machine-learning pipeline for real-time detection of gravitational waves from compact binary coalescences}},\ }\href@noop {} {\  (\bibinfo {year} {2024})},\ \Eprint {https://arxiv.org/abs/2403.18661} {arXiv:2403.18661 [gr-qc]} \BibitemShut {NoStop}%
\bibitem [{\citenamefont {McLeod}\ \emph {et~al.}(2024)\citenamefont {McLeod}, \citenamefont {Beveridge}, \citenamefont {Wen},\ and\ \citenamefont {Wicenec}}]{mcleod2024binaryneutronstarmerger}%
  \BibitemOpen
  \bibfield  {author} {\bibinfo {author} {\bibfnamefont {A.}~\bibnamefont {McLeod}}, \bibinfo {author} {\bibfnamefont {D.}~\bibnamefont {Beveridge}}, \bibinfo {author} {\bibfnamefont {L.}~\bibnamefont {Wen}},\ and\ \bibinfo {author} {\bibfnamefont {A.}~\bibnamefont {Wicenec}},\ }\href {https://arxiv.org/abs/2409.06266} {\bibinfo {title} {A binary neutron star merger search pipeline powered by deep learning}} (\bibinfo {year} {2024}),\ \Eprint {https://arxiv.org/abs/2409.06266} {arXiv:2409.06266 [gr-qc]} \BibitemShut {NoStop}%
\bibitem [{\citenamefont {Ormiston}\ \emph {et~al.}(2020)\citenamefont {Ormiston}, \citenamefont {Nguyen}, \citenamefont {Coughlin}, \citenamefont {Adhikari},\ and\ \citenamefont {Katsavounidis}}]{Ormiston:2020ele}%
  \BibitemOpen
  \bibfield  {author} {\bibinfo {author} {\bibfnamefont {R.}~\bibnamefont {Ormiston}}, \bibinfo {author} {\bibfnamefont {T.}~\bibnamefont {Nguyen}}, \bibinfo {author} {\bibfnamefont {M.}~\bibnamefont {Coughlin}}, \bibinfo {author} {\bibfnamefont {R.~X.}\ \bibnamefont {Adhikari}},\ and\ \bibinfo {author} {\bibfnamefont {E.}~\bibnamefont {Katsavounidis}},\ }\bibfield  {title} {\bibinfo {title} {{Noise Reduction in Gravitational-wave Data via Deep Learning}},\ }\href {https://doi.org/10.1103/PhysRevResearch.2.033066} {\bibfield  {journal} {\bibinfo  {journal} {Phys. Rev. Res.}\ }\textbf {\bibinfo {volume} {2}},\ \bibinfo {pages} {033066} (\bibinfo {year} {2020})},\ \Eprint {https://arxiv.org/abs/2005.06534} {arXiv:2005.06534 [astro-ph.IM]} \BibitemShut {NoStop}%
\bibitem [{\citenamefont {Bacon}\ \emph {et~al.}(2022)\citenamefont {Bacon}, \citenamefont {Trovato},\ and\ \citenamefont {Bejger}}]{Bacon:2022lsm}%
  \BibitemOpen
  \bibfield  {author} {\bibinfo {author} {\bibfnamefont {P.}~\bibnamefont {Bacon}}, \bibinfo {author} {\bibfnamefont {A.}~\bibnamefont {Trovato}},\ and\ \bibinfo {author} {\bibfnamefont {M.}~\bibnamefont {Bejger}},\ }\bibfield  {title} {\bibinfo {title} {{Denoising gravitational-wave signals from binary black holes with dilated convolutional autoencoder}},\ }\href@noop {} {\  (\bibinfo {year} {2022})},\ \Eprint {https://arxiv.org/abs/2205.13513} {arXiv:2205.13513 [gr-qc]} \BibitemShut {NoStop}%
\bibitem [{\citenamefont {Saleem}\ \emph {et~al.}(2023)\citenamefont {Saleem} \emph {et~al.}}]{Saleem:2023hcm}%
  \BibitemOpen
  \bibfield  {author} {\bibinfo {author} {\bibfnamefont {M.}~\bibnamefont {Saleem}} \emph {et~al.},\ }\bibfield  {title} {\bibinfo {title} {{Demonstration of Machine Learning-assisted real-time noise regression in gravitational wave detectors}},\ }\href@noop {} {\  (\bibinfo {year} {2023})},\ \Eprint {https://arxiv.org/abs/2306.11366} {arXiv:2306.11366 [gr-qc]} \BibitemShut {NoStop}%
\bibitem [{\citenamefont {Gunny}\ \emph {et~al.}(2022)\citenamefont {Gunny}, \citenamefont {Rankin}, \citenamefont {Krupa}, \citenamefont {Saleem}, \citenamefont {Nguyen}, \citenamefont {Coughlin}, \citenamefont {Harris}, \citenamefont {Katsavounidis}, \citenamefont {Timm},\ and\ \citenamefont {Holzman}}]{Gunny:2021gne}%
  \BibitemOpen
  \bibfield  {author} {\bibinfo {author} {\bibfnamefont {A.}~\bibnamefont {Gunny}}, \bibinfo {author} {\bibfnamefont {D.}~\bibnamefont {Rankin}}, \bibinfo {author} {\bibfnamefont {J.}~\bibnamefont {Krupa}}, \bibinfo {author} {\bibfnamefont {M.}~\bibnamefont {Saleem}}, \bibinfo {author} {\bibfnamefont {T.}~\bibnamefont {Nguyen}}, \bibinfo {author} {\bibfnamefont {M.}~\bibnamefont {Coughlin}}, \bibinfo {author} {\bibfnamefont {P.}~\bibnamefont {Harris}}, \bibinfo {author} {\bibfnamefont {E.}~\bibnamefont {Katsavounidis}}, \bibinfo {author} {\bibfnamefont {S.}~\bibnamefont {Timm}},\ and\ \bibinfo {author} {\bibfnamefont {B.}~\bibnamefont {Holzman}},\ }\bibfield  {title} {\bibinfo {title} {{Hardware-accelerated Inference for Real-Time Gravitational-Wave Astronomy}},\ }\href {https://doi.org/10.1038/s41550-022-01651-w} {\bibfield  {journal} {\bibinfo  {journal} {Nature Astron.}\ }\textbf {\bibinfo {volume} {6}},\ \bibinfo {pages} {529} (\bibinfo {year} {2022})},\ \Eprint {https://arxiv.org/abs/2108.12430}
  {arXiv:2108.12430 [gr-qc]} \BibitemShut {NoStop}%
\bibitem [{\citenamefont {Liao}\ and\ \citenamefont {Lin}(2021)}]{PhysRevD.103.124051}%
  \BibitemOpen
  \bibfield  {author} {\bibinfo {author} {\bibfnamefont {C.-H.}\ \bibnamefont {Liao}}\ and\ \bibinfo {author} {\bibfnamefont {F.-L.}\ \bibnamefont {Lin}},\ }\bibfield  {title} {\bibinfo {title} {Deep generative models of gravitational waveforms via conditional autoencoder},\ }\href {https://doi.org/10.1103/PhysRevD.103.124051} {\bibfield  {journal} {\bibinfo  {journal} {Phys. Rev. D}\ }\textbf {\bibinfo {volume} {103}},\ \bibinfo {pages} {124051} (\bibinfo {year} {2021})}\BibitemShut {NoStop}%
\bibitem [{\citenamefont {Chatterjee}\ \emph {et~al.}(2021)\citenamefont {Chatterjee}, \citenamefont {Wen}, \citenamefont {Diakogiannis},\ and\ \citenamefont {Vinsen}}]{Chatterjee_2021}%
  \BibitemOpen
  \bibfield  {author} {\bibinfo {author} {\bibfnamefont {C.}~\bibnamefont {Chatterjee}}, \bibinfo {author} {\bibfnamefont {L.}~\bibnamefont {Wen}}, \bibinfo {author} {\bibfnamefont {F.}~\bibnamefont {Diakogiannis}},\ and\ \bibinfo {author} {\bibfnamefont {K.}~\bibnamefont {Vinsen}},\ }\bibfield  {title} {\bibinfo {title} {Extraction of binary black hole gravitational wave signals from detector data using deep learning},\ }\bibfield  {journal} {\bibinfo  {journal} {Physical Review D}\ }\textbf {\bibinfo {volume} {104}},\ \href {https://doi.org/10.1103/physrevd.104.064046} {10.1103/physrevd.104.064046} (\bibinfo {year} {2021})\BibitemShut {NoStop}%
\bibitem [{\citenamefont {Sankarapandian}\ and\ \citenamefont {Kulis}(2021)}]{Sankarapandian:2021qun}%
  \BibitemOpen
  \bibfield  {author} {\bibinfo {author} {\bibfnamefont {S.}~\bibnamefont {Sankarapandian}}\ and\ \bibinfo {author} {\bibfnamefont {B.}~\bibnamefont {Kulis}},\ }\bibfield  {title} {\bibinfo {title} {{$\beta$-Annealed Variational Autoencoder for glitches}},\ }\href@noop {} {\  (\bibinfo {year} {2021})},\ \Eprint {https://arxiv.org/abs/2107.10667} {arXiv:2107.10667 [cs.LG]} \BibitemShut {NoStop}%
\bibitem [{\citenamefont {Morawski}\ \emph {et~al.}(2021)\citenamefont {Morawski} \emph {et~al.}}]{Morawski:2021kxv}%
  \BibitemOpen
  \bibfield  {author} {\bibinfo {author} {\bibfnamefont {F.}~\bibnamefont {Morawski}} \emph {et~al.},\ }\bibfield  {title} {\bibinfo {title} {Anomaly detection in gravitational waves data using convolutional autoencoders},\ }\href {http://iopscience.iop.org/article/10.1088/2632-2153/abf3d0} {\bibfield  {journal} {\bibinfo  {journal} {Machine Learning: Science and Technology}\ } (\bibinfo {year} {2021})}\BibitemShut {NoStop}%
\bibitem [{\citenamefont {Moreno}(2021)}]{eric_moreno_2021_5772814}%
  \BibitemOpen
  \bibfield  {author} {\bibinfo {author} {\bibfnamefont {E.}~\bibnamefont {Moreno}},\ }\href {https://doi.org/10.5281/zenodo.5772814} {\bibinfo {title} {{Source-Agnostic Gravitational-Wave Detection with Recurrent Autoencoders: BBH Dataset}}} (\bibinfo {year} {2021})\BibitemShut {NoStop}%
\bibitem [{\citenamefont {Verma}\ \emph {et~al.}(2024)\citenamefont {Verma}, \citenamefont {Reza}, \citenamefont {Gaur}, \citenamefont {Krishnaswamy},\ and\ \citenamefont {Caudill}}]{verma2024detectiongravitationalwavesignals}%
  \BibitemOpen
  \bibfield  {author} {\bibinfo {author} {\bibfnamefont {C.}~\bibnamefont {Verma}}, \bibinfo {author} {\bibfnamefont {A.}~\bibnamefont {Reza}}, \bibinfo {author} {\bibfnamefont {G.}~\bibnamefont {Gaur}}, \bibinfo {author} {\bibfnamefont {D.}~\bibnamefont {Krishnaswamy}},\ and\ \bibinfo {author} {\bibfnamefont {S.}~\bibnamefont {Caudill}},\ }\href {https://arxiv.org/abs/2206.12673} {\bibinfo {title} {Detection of gravitational wave signals from precessing binary black hole systems using convolutional neural network}} (\bibinfo {year} {2024}),\ \Eprint {https://arxiv.org/abs/2206.12673} {arXiv:2206.12673 [gr-qc]} \BibitemShut {NoStop}%
\bibitem [{\citenamefont {Baltus}\ \emph {et~al.}(2021)\citenamefont {Baltus}, \citenamefont {Janquart}, \citenamefont {Lopez}, \citenamefont {Reza}, \citenamefont {Caudill},\ and\ \citenamefont {Cudell}}]{PhysRevD.103.102003}%
  \BibitemOpen
  \bibfield  {author} {\bibinfo {author} {\bibfnamefont {G.}~\bibnamefont {Baltus}}, \bibinfo {author} {\bibfnamefont {J.}~\bibnamefont {Janquart}}, \bibinfo {author} {\bibfnamefont {M.}~\bibnamefont {Lopez}}, \bibinfo {author} {\bibfnamefont {A.}~\bibnamefont {Reza}}, \bibinfo {author} {\bibfnamefont {S.}~\bibnamefont {Caudill}},\ and\ \bibinfo {author} {\bibfnamefont {J.-R.}\ \bibnamefont {Cudell}},\ }\bibfield  {title} {\bibinfo {title} {Convolutional neural networks for the detection of the early inspiral of a gravitational-wave signal},\ }\href {https://doi.org/10.1103/PhysRevD.103.102003} {\bibfield  {journal} {\bibinfo  {journal} {Phys. Rev. D}\ }\textbf {\bibinfo {volume} {103}},\ \bibinfo {pages} {102003} (\bibinfo {year} {2021})}\BibitemShut {NoStop}%
\bibitem [{\citenamefont {Krastev}(2020)}]{Krastev_2020}%
  \BibitemOpen
  \bibfield  {author} {\bibinfo {author} {\bibfnamefont {P.~G.}\ \bibnamefont {Krastev}},\ }\bibfield  {title} {\bibinfo {title} {Real-time detection of gravitational waves from binary neutron stars using artificial neural networks},\ }\href {https://doi.org/10.1016/j.physletb.2020.135330} {\bibfield  {journal} {\bibinfo  {journal} {Physics Letters B}\ }\textbf {\bibinfo {volume} {803}},\ \bibinfo {pages} {135330} (\bibinfo {year} {2020})}\BibitemShut {NoStop}%
\bibitem [{\citenamefont {George}\ and\ \citenamefont {Huerta}(2018{\natexlab{c}})}]{PhysRevD.97.044039}%
  \BibitemOpen
  \bibfield  {author} {\bibinfo {author} {\bibfnamefont {D.}~\bibnamefont {George}}\ and\ \bibinfo {author} {\bibfnamefont {E.~A.}\ \bibnamefont {Huerta}},\ }\bibfield  {title} {\bibinfo {title} {Deep neural networks to enable real-time multimessenger astrophysics},\ }\href {https://doi.org/10.1103/PhysRevD.97.044039} {\bibfield  {journal} {\bibinfo  {journal} {Phys. Rev. D}\ }\textbf {\bibinfo {volume} {97}},\ \bibinfo {pages} {044039} (\bibinfo {year} {2018}{\natexlab{c}})}\BibitemShut {NoStop}%
\bibitem [{\citenamefont {Nousi}\ \emph {et~al.}(2023)\citenamefont {Nousi}, \citenamefont {Koloniari}, \citenamefont {Passalis}, \citenamefont {Iosif}, \citenamefont {Stergioulas},\ and\ \citenamefont {Tefas}}]{PhysRevD.108.024022}%
  \BibitemOpen
  \bibfield  {author} {\bibinfo {author} {\bibfnamefont {P.}~\bibnamefont {Nousi}}, \bibinfo {author} {\bibfnamefont {A.~E.}\ \bibnamefont {Koloniari}}, \bibinfo {author} {\bibfnamefont {N.}~\bibnamefont {Passalis}}, \bibinfo {author} {\bibfnamefont {P.}~\bibnamefont {Iosif}}, \bibinfo {author} {\bibfnamefont {N.}~\bibnamefont {Stergioulas}},\ and\ \bibinfo {author} {\bibfnamefont {A.}~\bibnamefont {Tefas}},\ }\bibfield  {title} {\bibinfo {title} {Deep residual networks for gravitational wave detection},\ }\href {https://doi.org/10.1103/PhysRevD.108.024022} {\bibfield  {journal} {\bibinfo  {journal} {Phys. Rev. D}\ }\textbf {\bibinfo {volume} {108}},\ \bibinfo {pages} {024022} (\bibinfo {year} {2023})}\BibitemShut {NoStop}%
\bibitem [{\citenamefont {Skliris}\ \emph {et~al.}(2024)\citenamefont {Skliris}, \citenamefont {Norman},\ and\ \citenamefont {Sutton}}]{skliris2024realtimedetectionunmodelledgravitationalwave}%
  \BibitemOpen
  \bibfield  {author} {\bibinfo {author} {\bibfnamefont {V.}~\bibnamefont {Skliris}}, \bibinfo {author} {\bibfnamefont {M.~R.~K.}\ \bibnamefont {Norman}},\ and\ \bibinfo {author} {\bibfnamefont {P.~J.}\ \bibnamefont {Sutton}},\ }\href {https://arxiv.org/abs/2009.14611} {\bibinfo {title} {Real-time detection of unmodelled gravitational-wave transients using convolutional neural networks}} (\bibinfo {year} {2024}),\ \Eprint {https://arxiv.org/abs/2009.14611} {arXiv:2009.14611 [astro-ph.IM]} \BibitemShut {NoStop}%
\bibitem [{\citenamefont {Raikman}\ \emph {et~al.}(2024)\citenamefont {Raikman} \emph {et~al.}}]{Raikman:2023ktu}%
  \BibitemOpen
  \bibfield  {author} {\bibinfo {author} {\bibfnamefont {R.}~\bibnamefont {Raikman}} \emph {et~al.},\ }\bibfield  {title} {\bibinfo {title} {{GWAK: gravitational-wave anomalous knowledge with recurrent autoencoders}},\ }\href {https://doi.org/10.1088/2632-2153/ad3a31} {\bibfield  {journal} {\bibinfo  {journal} {Mach. Learn. Sci. Tech.}\ }\textbf {\bibinfo {volume} {5}},\ \bibinfo {pages} {025020} (\bibinfo {year} {2024})},\ \Eprint {https://arxiv.org/abs/2309.11537} {arXiv:2309.11537 [astro-ph.IM]} \BibitemShut {NoStop}%
\bibitem [{\citenamefont {{Abbott}}\ \emph {et~al.}(2020)\citenamefont {{Abbott}} \emph {et~al.}}]{targeted_SN_O1-2}%
  \BibitemOpen
  \bibfield  {author} {\bibinfo {author} {\bibfnamefont {B.~P.}\ \bibnamefont {{Abbott}}} \emph {et~al.} (\bibinfo {collaboration} {LIGO Scientific Collaboration and Virgo Collaboration}),\ }\bibfield  {title} {\bibinfo {title} {{Optically targeted search for gravitational waves emitted by core-collapse supernovae during the first and second observing runs of advanced {LIGO} and advanced Virgo}},\ }\href {https://doi.org/10.1103/PhysRevD.101.084002} {\bibfield  {journal} {\bibinfo  {journal} {\prd}\ }\textbf {\bibinfo {volume} {101}},\ \bibinfo {eid} {084002} (\bibinfo {year} {2020})}\BibitemShut {NoStop}%
\bibitem [{\citenamefont {Abbott}\ \emph {et~al.}(2019{\natexlab{b}})\citenamefont {Abbott} \emph {et~al.}}]{O2magnetarbursts}%
  \BibitemOpen
  \bibfield  {author} {\bibinfo {author} {\bibfnamefont {B.~P.}\ \bibnamefont {Abbott}} \emph {et~al.} (\bibinfo {collaboration} {LIGO Scientific Collaboration and Virgo Collaboration}),\ }\bibfield  {title} {\bibinfo {title} {Search for transient gravitational-wave signals associated with magnetar bursts during advanced ligo’s second observing run},\ }\href {https://doi.org/10.3847/1538-4357/ab0e15} {\bibfield  {journal} {\bibinfo  {journal} {Astrophys. J.}\ }\textbf {\bibinfo {volume} {874}},\ \bibinfo {pages} {163} (\bibinfo {year} {2019}{\natexlab{b}})}\BibitemShut {NoStop}%
\bibitem [{\citenamefont {Abadie}\ \emph {et~al.}(2011)\citenamefont {Abadie} \emph {et~al.}}]{S6_NS}%
  \BibitemOpen
  \bibfield  {author} {\bibinfo {author} {\bibfnamefont {J.}~\bibnamefont {Abadie}} \emph {et~al.} (\bibinfo {collaboration} {LIGO Scientific Collaboration}),\ }\bibfield  {title} {\bibinfo {title} {{Search for gravitational waves associated with the August 2006 timing glitch of the Vela pulsar}},\ }\href {https://journals.aps.org/prd/abstract/10.1103/PhysRevD.83.042001} {\bibfield  {journal} {\bibinfo  {journal} {Phys. Rev. D}\ }\textbf {\bibinfo {volume} {83}},\ \bibinfo {pages} {042001} (\bibinfo {year} {2011})}\BibitemShut {NoStop}%
\bibitem [{\citenamefont {Ebersold}\ and\ \citenamefont {Tiwari}(2020)}]{o2_mem}%
  \BibitemOpen
  \bibfield  {author} {\bibinfo {author} {\bibfnamefont {M.}~\bibnamefont {Ebersold}}\ and\ \bibinfo {author} {\bibfnamefont {S.}~\bibnamefont {Tiwari}},\ }\bibfield  {title} {\bibinfo {title} {{Search for nonlinear memory from subsolar mass compact binary mergers}},\ }\href {https://doi.org/10.1103/PhysRevD.101.104041} {\bibfield  {journal} {\bibinfo  {journal} {Phys. Rev. D}\ }\textbf {\bibinfo {volume} {101}},\ \bibinfo {pages} {104041} (\bibinfo {year} {2020})}\BibitemShut {NoStop}%
\bibitem [{\citenamefont {Abbott}\ \emph {et~al.}(2021{\natexlab{c}})\citenamefont {Abbott} \emph {et~al.}}]{O3cosmicstring}%
  \BibitemOpen
  \bibfield  {author} {\bibinfo {author} {\bibfnamefont {R.}~\bibnamefont {Abbott}} \emph {et~al.} (\bibinfo {collaboration} {LIGO Scientific Collaboration, Virgo Collaboration and KAGRA Collaboration}),\ }\bibfield  {title} {\bibinfo {title} {{Constraints on cosmic strings using data from the third Advanced LIGO-Virgo observing run}},\ }\href {https://journals.aps.org/prl/abstract/10.1103/PhysRevLett.126.241102} {\bibfield  {journal} {\bibinfo  {journal} {Phys. Rev. Lett.}\ }\textbf {\bibinfo {volume} {126}},\ \bibinfo {pages} {241102} (\bibinfo {year} {2021}{\natexlab{c}})}\BibitemShut {NoStop}%
\bibitem [{\citenamefont {Abbott}\ \emph {et~al.}(2019{\natexlab{c}})\citenamefont {Abbott} \emph {et~al.}}]{allskyo2}%
  \BibitemOpen
  \bibfield  {author} {\bibinfo {author} {\bibfnamefont {B.~P.}\ \bibnamefont {Abbott}} \emph {et~al.} (\bibinfo {collaboration} {LIGO Scientific Collaboration, Virgo Collaboration}),\ }\bibfield  {title} {\bibinfo {title} {All-sky search for short gravitational-wave bursts in the second advanced {LIGO} and advanced virgo run},\ }\href {http://dx.doi.org/10.1103/PhysRevD.100.024017} {\bibfield  {journal} {\bibinfo  {journal} {Phys. Rev. D}\ }\textbf {\bibinfo {volume} {100}},\ \bibinfo {pages} {024017} (\bibinfo {year} {2019}{\natexlab{c}})}\BibitemShut {NoStop}%
\bibitem [{\citenamefont {{Abbott}}\ \emph {et~al.}(2021)\citenamefont {{Abbott}} \emph {et~al.}}]{O3IMBH}%
  \BibitemOpen
  \bibfield  {author} {\bibinfo {author} {\bibfnamefont {R.}~\bibnamefont {{Abbott}}} \emph {et~al.} (\bibinfo {collaboration} {LIGO Scientific Collaboration, Virgo Collaboration and KAGRA Collaboration}),\ }\bibfield  {title} {\bibinfo {title} {Intermediate mass black hole binary search in the third observing run of advanced {LIGO} and advanced virgo},\ }\href@noop {} {\bibfield  {journal} {\bibinfo  {journal} {{arXiv e-prints}}\ } (\bibinfo {year} {2021})},\ \Eprint {https://arxiv.org/abs/2105.15120} {arXiv:2105.15120 [astro-ph]} \BibitemShut {NoStop}%
\bibitem [{\citenamefont {The LIGO Scientific~Collaboration}\ and\ \citenamefont {the KAGRA~Collaboration}(2023)}]{Abbott_2023}%
  \BibitemOpen
  \bibfield  {author} {\bibinfo {author} {\bibfnamefont {t.~V.~C.}\ \bibnamefont {The LIGO Scientific~Collaboration}}\ and\ \bibinfo {author} {\bibnamefont {the KAGRA~Collaboration}},\ }\bibfield  {title} {\bibinfo {title} {Open data from the third observing run of {LIGO}, virgo, kagra, and geo},\ }\href {https://doi.org/10.3847/1538-4365/acdc9f} {\bibfield  {journal} {\bibinfo  {journal} {The Astrophysical Journal Supplement Series}\ }\textbf {\bibinfo {volume} {267}},\ \bibinfo {pages} {29} (\bibinfo {year} {2023})}\BibitemShut {NoStop}%
\bibitem [{\citenamefont {Buikema}\ \emph {et~al.}(2020)\citenamefont {Buikema} \emph {et~al.}}]{O3performance}%
  \BibitemOpen
  \bibfield  {author} {\bibinfo {author} {\bibfnamefont {A.}~\bibnamefont {Buikema}} \emph {et~al.},\ }\bibfield  {title} {\bibinfo {title} {Sensitivity and performance of the advanced {LIGO} detectors in the third observing run},\ }\href {https://doi.org/10.1103/PhysRevD.102.062003} {\bibfield  {journal} {\bibinfo  {journal} {Phys. Rev. D}\ }\textbf {\bibinfo {volume} {102}},\ \bibinfo {pages} {062003} (\bibinfo {year} {2020})}\BibitemShut {NoStop}%
\bibitem [{\citenamefont {Abbott}\ \emph {et~al.}(2016{\natexlab{b}})\citenamefont {Abbott} \emph {et~al.}}]{Abbott_2016}%
  \BibitemOpen
  \bibfield  {author} {\bibinfo {author} {\bibfnamefont {B.~P.}\ \bibnamefont {Abbott}} \emph {et~al.},\ }\bibfield  {title} {\bibinfo {title} {Characterization of transient noise in advanced {LIGO} relevant to gravitational wave signal gw150914},\ }\href {https://doi.org/10.1088/0264-9381/33/13/134001} {\bibfield  {journal} {\bibinfo  {journal} {Class. Quant. Grav.}\ }\textbf {\bibinfo {volume} {33}},\ \bibinfo {pages} {134001} (\bibinfo {year} {2016}{\natexlab{b}})}\BibitemShut {NoStop}%
\bibitem [{\citenamefont {Nguyen}\ \emph {et~al.}(2021)\citenamefont {Nguyen} \emph {et~al.}}]{envnoise}%
  \BibitemOpen
  \bibfield  {author} {\bibinfo {author} {\bibfnamefont {P.}~\bibnamefont {Nguyen}} \emph {et~al.},\ }\bibfield  {title} {\bibinfo {title} {Environmental noise in advanced {LIGO} detectors},\ }\href@noop {} {\bibfield  {journal} {\bibinfo  {journal} {arXiv e-prints}\ } (\bibinfo {year} {2021})},\ \Eprint {https://arxiv.org/abs/2101.09935} {arXiv:2101.09935 [astro-ph.IM]} \BibitemShut {NoStop}%
\bibitem [{\citenamefont {Effler}\ \emph {et~al.}(2015)\citenamefont {Effler}, \citenamefont {Schofield}, \citenamefont {Frolov}, \citenamefont {Gonz{\'{a}}lez}, \citenamefont {Kawabe}, \citenamefont {Smith}, \citenamefont {Birch},\ and\ \citenamefont {McCarthy}}]{PEM}%
  \BibitemOpen
  \bibfield  {author} {\bibinfo {author} {\bibfnamefont {A.}~\bibnamefont {Effler}}, \bibinfo {author} {\bibfnamefont {R.~M.~S.}\ \bibnamefont {Schofield}}, \bibinfo {author} {\bibfnamefont {V.~V.}\ \bibnamefont {Frolov}}, \bibinfo {author} {\bibfnamefont {G.}~\bibnamefont {Gonz{\'{a}}lez}}, \bibinfo {author} {\bibfnamefont {K.}~\bibnamefont {Kawabe}}, \bibinfo {author} {\bibfnamefont {J.~R.}\ \bibnamefont {Smith}}, \bibinfo {author} {\bibfnamefont {J.}~\bibnamefont {Birch}},\ and\ \bibinfo {author} {\bibfnamefont {R.}~\bibnamefont {McCarthy}},\ }\bibfield  {title} {\bibinfo {title} {Environmental influences on the {LIGO} gravitational wave detectors during the 6th science run},\ }\href {https://doi.org/10.1088/0264-9381/32/3/035017} {\bibfield  {journal} {\bibinfo  {journal} {Class. Quant. Grav.}\ }\textbf {\bibinfo {volume} {32}},\ \bibinfo {pages} {035017} (\bibinfo {year} {2015})}\BibitemShut {NoStop}%
\bibitem [{\citenamefont {Abbott}\ \emph {et~al.}(2018)\citenamefont {Abbott} \emph {et~al.}}]{dqmitigation}%
  \BibitemOpen
  \bibfield  {author} {\bibinfo {author} {\bibfnamefont {B.~P.}\ \bibnamefont {Abbott}} \emph {et~al.} (\bibinfo {collaboration} {LIGO Scientific Collaboration and Virgo Collaboration}),\ }\bibfield  {title} {\bibinfo {title} {Effects of data quality vetoes on a search for compact binary coalescences in advanced ligo's first observing run},\ }\href {https://iopscience.iop.org/article/10.1088/1361-6382/aaaafa} {\bibfield  {journal} {\bibinfo  {journal} {Class. Quant. Grav.}\ }\textbf {\bibinfo {volume} {35}},\ \bibinfo {pages} {065010} (\bibinfo {year} {2018})}\BibitemShut {NoStop}%
\bibitem [{\citenamefont {Aasi}\ \emph {et~al.}(2015)\citenamefont {Aasi} \emph {et~al.}}]{TheLIGOScientific:2014jea}%
  \BibitemOpen
  \bibfield  {author} {\bibinfo {author} {\bibfnamefont {J.}~\bibnamefont {Aasi}} \emph {et~al.} (\bibinfo {collaboration} {LIGO Scientific}),\ }\bibfield  {title} {\bibinfo {title} {{Advanced {LIGO}}},\ }\href {https://doi.org/10.1088/0264-9381/32/7/074001} {\bibfield  {journal} {\bibinfo  {journal} {Class. Quant. Grav.}\ }\textbf {\bibinfo {volume} {32}},\ \bibinfo {pages} {074001} (\bibinfo {year} {2015})},\ \Eprint {https://arxiv.org/abs/1411.4547} {arXiv:1411.4547 [gr-qc]} \BibitemShut {NoStop}%
\bibitem [{\citenamefont {Robinet}\ \emph {et~al.}(2020)\citenamefont {Robinet}, \citenamefont {Arnaud}, \citenamefont {Leroy}, \citenamefont {Lundgren}, \citenamefont {Macleod},\ and\ \citenamefont {McIver}}]{Robinet:2020lbf}%
  \BibitemOpen
  \bibfield  {author} {\bibinfo {author} {\bibfnamefont {F.}~\bibnamefont {Robinet}}, \bibinfo {author} {\bibfnamefont {N.}~\bibnamefont {Arnaud}}, \bibinfo {author} {\bibfnamefont {N.}~\bibnamefont {Leroy}}, \bibinfo {author} {\bibfnamefont {A.}~\bibnamefont {Lundgren}}, \bibinfo {author} {\bibfnamefont {D.}~\bibnamefont {Macleod}},\ and\ \bibinfo {author} {\bibfnamefont {J.}~\bibnamefont {McIver}},\ }\bibfield  {title} {\bibinfo {title} {{Omicron: a tool to characterize transient noise in gravitational-wave detectors}},\ }\href {https://doi.org/10.1016/j.softx.2020.100620} {\bibfield  {journal} {\bibinfo  {journal} {SoftwareX}\ }\textbf {\bibinfo {volume} {12}},\ \bibinfo {pages} {100620} (\bibinfo {year} {2020})},\ \Eprint {https://arxiv.org/abs/2007.11374} {arXiv:2007.11374 [astro-ph.IM]} \BibitemShut {NoStop}%
\bibitem [{\citenamefont {Nitz}\ \emph {et~al.}(2020)\citenamefont {Nitz} \emph {et~al.}}]{alex_nitz_2020_3993665}%
  \BibitemOpen
  \bibfield  {author} {\bibinfo {author} {\bibfnamefont {A.}~\bibnamefont {Nitz}} \emph {et~al.},\ }\href {https://doi.org/10.5281/zenodo.3993665} {\bibinfo {title} {gwastro/pycbc: Pycbc release v1.16.9}} (\bibinfo {year} {2020})\BibitemShut {NoStop}%
\end{thebibliography}%

\end{document}